\documentstyle[12pt]{article}
\catcode`\@=11                                   
\@addtoreset{equation}{section}                  
\@addtoreset{footnote}{section}                  
\@addtoreset{footnote}{subsection}               
\catcode`\@=12                                   
\newtheorem{proposition}{Proposition}
\newtheorem{lemma}{Lemma}
\newtheorem{theorem}{Theorem}
\newenvironment{remark}%
  {\bigskip\noindent{\it Remark}\ }%
  {\bigskip}
\newenvironment{proof}%
  {\bigskip\noindent{\it Proof}\ }%
  {$\Box$\bigskip}
  {\bigskip\noindent{\large\bf Acknowledgement}\medskip\par\noindent}%
  {\bigskip}
\newenvironment{note-added}%
  {\bigskip\noindent{\large\bf Note added}\medskip\par\noindent}%
  {\bigskip}
\newcommand{\beqn}{\begin{equation}}
\newcommand{\eeqn}{\end{equation}}
\newcommand{\beqnarray}{\begin{eqnarray}}
\newcommand{\eeqnarray}{\end{eqnarray}}
\newcommand{\rd}{\partial}
\newcommand{\dfrac}[2]{ \frac{\displaystyle #1}{\displaystyle #2} }
\newcommand{\res}{\;\mathop{\mbox{\rm res}}}
\newcommand{\tr}{\;\mathop{\mbox{\rm tr}}\nolimits}
\newcommand{\Dbar}{\bar{D}}
\newcommand{\cA}{{\cal A}}
\newcommand{\cB}{{\cal B}}
\newcommand{\cL}{{\cal L}}
\newcommand{\cLbar}{\bar{\cL}}
\newcommand{\cM}{{\cal M}}
\newcommand{\cMbar}{\bar{\cM}}
\newcommand{\cU}{{\cal U}}
\newcommand{\cUbar}{\bar{\cU}}
\newcommand{\cV}{{\cal V}}
\newcommand{\cVbar}{\bar{\cV}}
\newcommand{\cP}{{\cal P}}
\newcommand{\cQ}{{\cal Q}}
\newcommand{\cR}{{\cal R}}

\newcommand{\gln}{\mbox{\rm gl}}

\newcommand{\Poisson}{\mbox{\rm Poisson}}
\newcommand{\Moyal}{\mbox{\rm Moyal}}
\begin{document}

\begin{flushright}
\baselineskip=12pt
KUCP-0068\\
hep-th/9407098\\
July 1994
\end{flushright}

\begin{center}
\LARGE
Symmetries and tau function of \\
higher dimensional\\
dispersionless integrable hierarchies\\
\bigskip
\Large
    Kanehisa Takasaki \\
\normalsize\it
    Department of Fundamental Sciences\\
    Faculty of Integrated Human Studies, Kyoto University\\
    Yoshida-Nihonmatsu-cho, Sakyo-ku, Kyoto 606, Japan\\
\rm
    E-mail: takasaki @ jpnyitp.yukawa.kyoto-u.ac.jp\\
\end{center}
\bigskip
\begin{abstract}
\noindent
A higher dimensional analogue of the dispersionless KP hierarchy
is introduced.  In addition to the two-dimensional ``phase space''
variables $(k,x)$ of the dispersionless KP hierarchy, this hierarchy
has extra spatial dimensions compactified to a two (or any even)
dimensional torus. Integrability of this hierarchy and the existence
of an infinite dimensional of ``additional symmetries'' are ensured
by an underlying twistor theoretical structure (or a nonlinear
Riemann-Hilbert problem).  An analogue of the tau function, whose
logarithm gives the $F$ function (``free energy'' or ``prepotential''
in the contest of matrix models and topological conformal field
theories), is constructed. The infinite dimensional symmetries can
be extended to this tau (or $F$) function. The extended symmetries,
just like those of the dispersionless KP hierarchy, obey an anomalous
commutation relations.
\end{abstract}
\newpage

\section{Introduction}

Recently we proposed a new family of higher dimensional integrable
hierarchies \cite{bib:T-MoyalKP}. These hierarchies can be divided
into two types. One is formulated in terms of pseudo-differential
operators with Moyal algebraic coefficients, and considered to
be a higher dimensional analogue of the ordinary KP hierarchy.
The other is obtained as classical limit ($\hbar \to 0$) of these
hierarchies. In this limit, the Moyal algebra turns into a Poisson
algebra, and the hierarchies are formulated in terms of Poisson
brackets rather than commutators. Because of this, the latter
hierarchies are viewed as a higher dimensional analogue of the
dispersionless KP hierarchy.  The manifold carrying the Moyal or
Poisson algebra emerges as ``extra spatial dimensions'';
this is a reason that we call these hierarchies
``higher dimensional''.

These higher dimensional integrable hierarchies naturally inherit
various characteristics of ordinary integrable hierarchies.  In
particular, one can develop a Lax formalism and a Riemann-Hilbert
problem method in an almost parallel way. The notion of ``tau
functions'' for these hierarchies, however, has remained obscure.
Since tau functions are the most fundamental quantities in low
dimensional integrable hierarchies
\cite{bib:Sato-Sato,bib:DJKM,bib:Segal-Wilson},
extending the theory of tau functions to these new hierarchies
will be a crucial issue for better understanding of higher
dimensional integrable hierarchies.

As far as we know, no essentially higher dimensional extension
of tau functions has been constructed.  Those that are already
proposed will be presumably interpreted as the tau function of
a multi-component KP (or Toda) hierarchy.  Another possible
candidate of a higher dimensional tau function is the Plebanski
function (K\"ahler potential) $\Omega$ of self-dual Einstein
gravity \cite{bib:Plebanski}. If, however, one attempts to
carefully compare $\Omega$ with, for instance, the tau function
$\tau_{dKP}$ of the dispersionless KP hierarchy (and related
integrable hierarchies) \cite{bib:Krichever,bib:Dubrovin,bib:TT-dKP},
it will turn out that $\Omega$ corresponds to the spatial
logarithmic derivative $\rd_x \log\tau_{dKP}$ rather than to
$\tau_{dKP}$ itself. This difference is also reflected in their
behavior under ``additional symmetries'' of these systems.
Additional symmetries of these systems both carry a Lie algebraic
structure related to two-dimensional Poisson algebras, but whereas
the Lie algebraic structure of $\Omega$ is the Poisson algebra
itself \cite{bib:Boyer-Plebanski,bib:T-SDG,bib:Park-PLB},
the Lie algebra emerging in $\tau_{dKP}$ is a central extension
of the Poisson algebra \cite{bib:TT-dKP}.  The presence of such a
central extension is a common characteristic of tau functions of
low dimensional integrable integrable hierarchies. We expect that
a higher dimensional tau function, if any, should possess a similar
property.

As the first step towards such a higher dimensional theory of
tau functions, we propose an analogue of $\tau_{dKP}$ for a
higher dimensional dispersionless integrable hierarchy. More
precisely, it is the $F$ function
\beqn
    F = \log\tau_{dKP}
\eeqn
rather
than the tau function $\tau_{dKP}$ itself that plays a fundamental
role in the dispersionless KP hierarchy; $F$ is also called the
``free energy'' or ``prepotential'' in the context of matrix models
and topological conformal field theories. We shall construct a
higher dimensional analogue of this $F$ function.

To obtain such a higher dimensional $F$ function, we have to
slightly modify our previous construction \cite{bib:T-MoyalKP} of
higher dimensional integrable hierarchies.  Our higher dimensional
integrable hierarchies are characterized by a Moyal or Poisson
algebra on extra spatial dimensions, which (for simplicity) we
assume to be a two-dimensional surface $\Sigma$. In the previous
models (planar models), $\Sigma$ is a two-dimensional plane.
This is not suited for constructing an $F$ function.  We now
consider a model (toroidal model) for which $\Sigma$ is a
two-dimensional torus $T^2$. Unlike the planer case, the toroidal
Poisson algebra has the linear functional
\beqn
    a \mapsto
      \int_{T^2} \dfrac{d\theta_1 d\theta_2}{(2\pi)^2}
        a(\theta_1,\theta_2),
\eeqn
where $(\theta_1,\theta_2)$ are angle variables on the torus $T^2$.
This linear functional on the Poisson algebra is, in a sense, a
remnant of ``trace'', because it satisfies the identity
\beqn
    \int_{T^2} \dfrac{d\theta_1 d\theta_2}{(2\pi)^2}
      \{ a, b \}_\theta
        = 0,
\eeqn
where $\{\quad,\quad\}_\theta$ denotes the standard Poisson bracket
\beqn
    \{ a, b \}_\theta
      = \dfrac{\rd a}{\rd \theta_1} \dfrac{\rd b}{\rd \theta_2}
        - \dfrac{\rd a}{\rd \theta_2} \dfrac{\rd b}{\rd \theta_1}.
\eeqn
Actually, the same identity is also satisfied by the Moyal bracket
\beqn
  \{ a, b \}_{\hbar,\theta}
  = \frac{2}{\hbar}
     \left.
       \sin \left[
        \dfrac{\hbar}{2} \left(
           \frac{\rd^2}{\rd \theta_1 \rd \theta_2'}
          -\frac{\rd^2}{\rd \theta_2 \rd \theta_1'} \right)
       \right] a(\theta_1,\theta_2) b(\theta_1',\theta_2')
     \right|_{\theta_1'=\theta_1,\theta_2'=\theta_2},
\eeqn
and the above linear functional may be understood as a trace
functional on the toroidal Moyal algebra \cite{bib:Moyal}.
The above identity for the Poisson bracket is its classical
limit. It is this trace functional that allows us to define a
higher dimensional $F$ function.  In principle, one should be
abel to deal with more general models with such a trace
functional. The reason that we now focus on the case of
$\Sigma = T^2$ is rather technical; in that case, the structure
of the integrable hierarchy is almost the same as the planar
model, so that the construction of the $F$ function relatively
simplifies.

The role of such a ``trace functional'' will become most clear
from the point of view of ``large-$N$ limit''.  As stressed in
the previous paper \cite{bib:T-MoyalKP}, our higher dimensional
integrable hierarchies may be thought of as ``large-$N$ limit''
of $N$-component KP hierarchies.  The origin of the extra spatial
dimensions lies in physicists' observation \cite{bib:large-N}
that the Lie algebra of $N \times N$ matrices turns into a Moyal
or Poisson algebra on a two-dimensional surface $\Sigma$ as
$N \to \infty$:
\beqn
  \gln(N) \mathop{\longrightarrow}_{N \to \infty}
  \left\{ \begin{array}{l}
            \Poisson(\Sigma) \\
            \Moyal(\Sigma)
         \end{array}
  \right.
\eeqn
This observation was a clue to Park's interpretation of
self-dual gravity as large-$N$ limit of two-dimensional
sigma models \cite{bib:Park-PLB} and its Moyal algebraic
deformations \cite{bib:Strachan}; we applied the same
idea to the $N$-component KP hierarchy. The $N$-component
KP hierarchy is formulated in terms of pseudo-differential
operators with $\gln(N)$ coefficients. In large-$N$ limit,
accordingly, those pseudo-differential operators will be
replaced by corresponding quantities (i.e., pseudo-differential
operators or phase space functions) with $Poisson(\Sigma)$
or $\Moyal(\Sigma)$ coefficients --- this heuristic
consideration lies in the heart of our previous work.

Since the notion of tau functions is extended to the
$N$-component KP hierarchy
\cite{bib:Sato-Sato,bib:DJKM,bib:Segal-Wilson}, it is
natural to attempt to consider large-$N$ limit of the tau
function of the $N$-component KP hierarchy. Actually,
life is not so simple, because the usual definition of
a tau function as an infinite determinant
\cite{bib:Sato-Sato,bib:Segal-Wilson} or the vacuum
expectation value of a free fermion theory \cite{bib:DJKM}
is (at least for the moment) not suited for attain this limit.

To overcome this difficulty, we rather resort to an older
version of the theory of tau functions developed in the
context of ``monodromy preserving deformations''
\cite{bib:Miwa-Jimbo-Ueno}.  Tau functions are defined
therein as a potential of a closed 1-form,
\beqn
    d\log\tau = \alpha = \sum \alpha_n dt_n,
\eeqn
and the coefficients on the right hand side are given by a trace
\beqn
    \alpha_n = \tr_{N \times N} A_n
\eeqn
of $\gln(N)$-valued functions that are comprised of Laurent
coefficients of a particular solution in an associated linear
problem. This is also the case for the $N$-component KP hierarchy
\cite{bib:DJKM}. In large-$N$ limit, the trace $\tr_{N \times N}$
should turn into the aforementioned trace functional on
$\Poisson(\Sigma)$ or $\Moyal(\Sigma)$. More precisely, we have
to ``renormalize'' $\tr_{N \times N}$ by the factor $1/N$:
\beqn
    \frac{1}{N} \tr_{N \times N} A
    \mathop{\rightarrow}_{N \to \infty}
    \tr_\Sigma a =
      \int_{\Sigma} \dfrac{d\theta_1 d\theta_2}{(2\pi)^2}
        a(\theta_1,\theta_2).
\eeqn
Combining this relation with the above defining equation of
$\log\tau$, one will be naturally led to imagine that a higher
dimensional tau function (or $F$ function) is defined in a
similar way as above:
\beqn
    d\log\tau = \tr_\Sigma(\ldots)
    \quad
    \Bigl(\mbox{or}\quad  dF = \tr_\Sigma(\ldots)  \Bigr).
\eeqn
We shall indeed define an $F$ function of the toroidal model
as such a potential, though the 1-form itself is obtained by
guess work (in analogy from the corresponding formula to the
dispersionless KP hierarchy) rather than literally chasing
this large-$N$ limit (which seems technically difficult).

To confirm that our definition of a higher dimensional $F$ function
is a correct one, we construct an infinite dimensional symmetries of
the hierarchy, and examine their algebraic properties concerning $F$.
The above ``trace functional'' enters into the construction of
symmetries on $F$, too, so that the planar model has to be ruled
out again.  Apart from this new ingredient, the construction of
symmetries is very similar (though considerably complicated) to
the case of the dispersionless KP hierarchy. As expected, those
symmetries obey anomalous commutation relations and give a central
extension of a four-dimensional Poisson algebra.

This paper is organized as follows. Sections 2 and 3 are rather
of preliminary nature.  In Section 2, we collect basic results
from our previous papers \cite{bib:TT-dKP,bib:TT-qcKP} and
review \cite{bib:TT-review}. These results are presented as a
prototype of the contents of subsequent sections. In Section 4,
we introduce our new hierarchy and show an underlying twistor
theoretical structure ensuring integrability of the hierarchy.
Sections 4 and 5 are devoted to presenting our main results.
In Section 4, we construct additional symmetries in the language
of the Lax formalism, and show that their commutation relations
obey an underlying Poisson algebraic structure. In Section 5, we
define the $F$ function and consider the symmetries of the
previous section in the language of this $F$ function.
Commutation relations of the extended symmetries are calculated
and shown to include an anomalous central term.  Section 6 is for
concluding remarks and discussions.

\section{Dispersionless KP hierarchy}

\subsection{Extended Lax formalism}

The extended Lax formalism of the dispersionless KP hierarchy
consists of the following three equations.
\beqnarray
  \dfrac{\rd\cL}{\rd t_n} &=& \{ \cB_n, \cL \}_{kx},
                                     \nonumber         \\
  \dfrac{\rd\cM}{\rd t_n} &=& \{ \cB_n, \cM \}_{kx},
                                     \label{eq:dKPLax} \\
  \{ \cL, \cM \}_{kx}     &=& 1.
                                     \label{eq:dKPCanRel}
\eeqnarray

The first equation of (\ref{eq:dKPLax}) corresponds to the
usual Lax representation of the KP hierarchy with an infinite
number of time variables $t = (t_1, t_2, ...)$. The dispersionless
analogue $\cL$ of the Lax operator $L$ is a Laurent series
of the form
\beqn
  \cL = k + \sum_{n=1}^\infty g_{n+1}(t,x) k^{-n},
\eeqn
and $\{\quad,\quad\}_{kx}$ stands for a Poisson bracket in
the two-dimensional ``phase space'' $(k,x)$,
\beqn
    \{ A, B \}_{kx}
  = \dfrac{\rd A}{\rd k} \dfrac{\rd B}{\rd x}
   -\dfrac{\rd A}{\rd x} \dfrac{\rd B}{\rd k}.
\eeqn
The dispersionless analogue $\cB_n$ of the Zakharov-Shabat
operators $B_n$ are given by
\beqn
    \cB_n = \left( \cL^n \right)_{\ge 0},
\eeqn
and obey the zero-curvature equations
\beqn
    \dfrac{\rd \cB_m}{\rd t_n} - \dfrac{\rd \cB_n}{\rd t_m}
    + \{ \cB_m, \cB_n \}_{kx} = 0.
\eeqn
Here $(\quad)_{\ge 0}$ denotes the projection onto the
polynomial part of Laurent series in $k$. Similarly, we shall
use $(\quad)_{\le -1}$ for the complementary projection (i.e.,
strictly negative power part):
\beqnarray
    \left( \sum a_n k^n \right)_{\ge 0} &=& \sum_{n \ge 0} a_n k^n,
                                                        \nonumber \\
    \left( \sum a_n k^n \right)_{\le -1} &=& \sum_{n \le -1} a_n k^n.
\eeqnarray

The status of the variable $t_1$ is special: Since $\cB_1 = k$,
the $t_1$-dependence of all quantities coincides with the
$x$-dependence,
\beqn
    \dfrac{\rd \cL}{\rd t_1} = \dfrac{\rd \cL}{\rd x}, \quad
    \dfrac{\rd \cM}{\rd t_1} = \dfrac{\rd \cM}{\rd x},
\eeqn
therefore all quantities of the hierarchy depend on $t_1$ and
$x$ only through the linear combination $t_1 + x$. We could have
accordingly put $t_1 = x$, but we rather distinguish
between them so as to clarify their roles as time and space
variables.

The second equation of (\ref{eq:dKPLax}) and the canonical
relation (\ref{eq:dKPCanRel}) characterize the dispersionless
analogue $\cM$ of the Orlov-Shulman operator $M$
\cite{bib:Orlov-Shulman}:
\beqn
    \cM = \sum_{n=1}^\infty n t_n \cL^{n-1} + x
        + \sum_{n=1}^\infty h_n(t,x) \cL^{-n-1}.
\eeqn
Such an extended Lax formalism of the KP hierarchy,
as pointed out by Orlov, Shulman and Grinevich
\cite{bib:Orlov-Shulman,bib:Grinevich-Orlov},
provides a very useful framework for describing
``additional symmetries'' of the KP hierarchy.
This is also the case of the dispersionless KP hierarchy.
Furthermore, the extended Lax representation of the
dispersionless case can also be rewritten into a 2-form
equation of the form
\beqn
    d\cL \wedge d\cM
      = dk \wedge dx + \sum_{n=1}^\infty d\cB_n \wedge dt_n,
                                           \label{eq:dKP2-form}
\eeqn
where ``$d$'' here stands for total differential in $(k,t,x)$.
This 2-form equation is a clue for applying Penrose's twistor
theoretical approach to the self-dual Einstein equation
(``nonlinear graviton construction'') \cite{bib:twistor}.
The integrability of the system and the existence of an
infinite dimensional symmetries are a consequence of this
twistor theoretical structure.

\subsection{Additional symmetries}

By ``symmetries'', we mean infinitesimal symmetries. Such
symmetries can be most clearly formulated in the abstract
language of differential rings.
\footnote{A differential ring is a ring $\cR$ with a set of
abstract derivations $\{\rd_i\}$, i.e., linear maps $\cR \to \cR$
obeying the Leibniz rule $\rd_i(ab) = (\rd_i a) b + a (\rd_i b)$,
$a,b \in \cR$.}  We now introduce two fundamental differential rings.

Let $\cR_0$ be the differential ring generated by $g_n$ and $h_n$.
As a commutative ring, $\cR_0$ has generators $g_n^{(\alpha)}$ and
$h_n^{(\alpha)}$, $\alpha = 0,1,\ldots$, which represent $\alpha$-th
derivatives of $g_n$ and $h_n$. We consider them as abstract symbols
rather than actual functions. These generators are subject to an
infinite set of algebraic equations to be derived from canonical
relation (\ref{eq:dKPCanRel}). This ring is further equipped with
abstract derivations $\rd$ and $\rd_n$ that represent $\rd/\rd x$
and $\rd/\rd t_n$. They are to be defined in an obvious way as:
\beqnarray
    \rd g_n^{(\alpha)} = g_n^{(\alpha+1)}, &\quad&
    \rd h_n^{(\alpha)} = h_n^{(\alpha+1)},
                                                    \nonumber \\
    \rd_n g_n^{(\alpha)} = \ldots,         &\quad&
    \rd_n h_n^{(\alpha)} = \ldots,
\eeqnarray
where the right hand side of the last two equations are the same
as the right hand side of the equations to be derived from
(\ref{eq:dKPLax}). Explicit forms are however irrelevant here.
The dispersionless KP hierarchy is thus encoded into the abstract
structure of the differential ring $\cR_0$.

Let $\cR$ be the extension of this differential ring by variables
$t_n$ and $x$.  For definiteness, we may take the ring of formal
power series: $\cR = \cR_0[[t,x]]$.  The abstract derivations $\rd$
and $\rd_n$ can be uniquely extended onto this larger ring by the
obvious rules
\beqnarray
    \rd x = 1,    && \rd t_m = 0,    \nonumber \\
    \rd_n x  = 0, && \rd_n t_m = \delta_{nm}.
\eeqnarray

We need such an extended ring in order to define the notion of
``additional symmetries''. By definition, an additional symmetry
(which, from now on, we simply call a symmetry) is a linear map
$\delta: \cR \to \cR$ that commute with all $\rd$ and $\rd_n$:
\beqn
    [ \delta, \rd ] = 0, \quad [ \delta, \rd_n ] = 0.
                                           \label{eq:SymmDef}
\eeqn
Actually, we only consider a more restricted class of symmetries,
i.e., symmetries that leave $x$ and $t_n$ invariant:
\beqn
    \delta x = 0, \quad \delta t_n = 0.
                                           \label{eq:SymmInner}
\eeqn
Such a symmetry represents an ``inner'' symmetry that commute with
spatial and time translations.

This abstract formulation of symmetries can be translated into
a more down-to-earth language.  Let us consider an infinitesimal
transformation $\cL \to \cL + \epsilon \delta\cL$,
$\cM \to \cM + \epsilon \delta\cM$ with an infinitesimal
parameter $\epsilon$, and look for a condition under which
the extended Lax equations are satisfied to the first order
of $\epsilon$, i.e.,
\beqn
    \dfrac{\rd}{\rd t_n}(\cL + \epsilon\cL)
           = \{ \cB_n + \epsilon \cB_n, \cL + \epsilon \cL \}_{kx}
             + O(\epsilon^2),
    \quad \mbox{etc.,}
\eeqn
where $\cB_n \to \cB_n + \epsilon \delta\cB_n$ is understood to
be the transformation induced by the relation between $\cB_n$ and
$\cL$. Here and in what follows, we again use the explicit notations
$\rd/\rd x$ and $\rd/\rd t_n$ rather than the abstract ones $\rd$
and $\rd_n$. Picking out terms of first order in $\epsilon$, we
obtain the following conditions for $\delta\cL$ and $\delta\cM$
to become a symmetry of the dispersionless KP hierarchy:
\beqnarray
    \dfrac{\rd}{\rd t_n} \delta\cL
      &=&   \{ \delta\cB_n, \cL \}_{kx}
          + \{ \cB_n, \delta\cL \}_{kx},
                                          \nonumber \\
    \dfrac{\rd}{\rd t_n} \delta\cM
      &=&   \{ \delta\cB_n, \cM \}_{kx}
          + \{ \cB_n, \delta\cM \}_{kx},
                                          \nonumber \\
    0 &=& \{ \delta\cL, \cM \}_{kx} + \{ \cL, \delta\cM \}_{kx}.
                                          \label{eq:dKPSymmCond1}
\eeqnarray
Recalling that $\delta$ is a derivation on $\cR$ (i.e., satisfies
the Leibniz rule), these conditions can be written more simply:
\beqnarray
    \dfrac{\rd}{\rd t_n} \delta\cL
      &=& \delta \left( \{ \cB_n, \cL \}_{kx} \right)
       =  \delta \dfrac{\rd \cL}{\rd t_n},
                                          \nonumber \\
    \dfrac{\rd}{\rd t_n} \delta\cM
      &=& \delta \left( \{ \cB_n, \cM \}_{kx} \right)
       =  \delta \dfrac{\rd \cM}{\rd t_n},
                                          \nonumber \\
    0 &=& \delta \{ \cL, \cM \}_{kx}.
                                          \label{eq:dKPSymmCond2}
\eeqnarray
The first and second equations of the last expression simply
mean that $\delta$ and $\rd/\rd_n$ commute. Since $\rd/\rd x$
is effectively the same as $\rd/\rd t_1$, $\delta$ also commutes
with $\rd/\rd x$. Thus we can reproduce (\ref{eq:SymmDef}).

Symmetries satisfying these conditions are explicitly constructed
as follows. Given an arbitrary Laurent series
\beqn
    \cA = \sum_{i=-\infty}^\infty \sum_{j=0}^\infty
          a_{ij} \lambda^i \mu^j,
\eeqn
one can define a derivation $\delta_\cA$ on $\cR$ by the
conditions
\beqnarray
    \delta_\cA \cL &=& \{ \cA(\cL,\cM)_{\le -1}, \cL \}_{kx},
                                              \nonumber \\
    \delta_\cA \cM &=& \{ \cA(\cL,\cM)_{\le -1}, \cM \}_{kx},
                                              \nonumber \\
    \delta_\cA x &=& \delta_\cA t_n = 0.
                                         \label{eq:dKPSymm}
\eeqnarray
This additional derivation $\delta_\cA$ turns out to satisfy
the above conditions to satisfy (\ref{eq:dKPSymmCond2}).
Furthermore, these symmetries obey the commutation relations
\beqn
    [ \delta_\cA, \delta_\cB ] = \delta_{ \{\cA,\cB\}_{\lambda\mu} },
                                             \label{eq:dKPCommRel}
\eeqn
where
\beqn
    \{ \cA, \cB \}_{\lambda\mu}
    =   \dfrac{\rd\cA}{\rd\lambda} \dfrac{\rd\cB}{\rd\mu}
      - \dfrac{\rd\cA}{\rd\mu} \dfrac{\rd\cB}{\rd\lambda}.
\eeqn
This the linear map $\cA \mapsto \delta_\cA$ gives a homomorphism
of a two-dimensional Poisson algebra into the algebra of symmetries
on $\cR$.

As already mentioned, the origin of these symmetries lies in the
twistor theoretical structure (and an underlying Riemann-Hilbert
problem). This has been discussed in our earlier work on the
dispersionless KP (and Toda) hierarchy \cite{bib:TT-dKP,bib:TT-dToda}
and its prototype in the self-dual Einstein equation
\cite{bib:Boyer-Plebanski,bib:T-SDG,bib:Park-PLB}.  These symmetries
can also be reproduced as a quasi-classical limit of $W_{1+\infty}$
symmetries of the KP hierarchy \cite{bib:TT-qcKP}.

\subsection{F function and symmetries}

The $F$ function of the dispersionless KP hierarchy was first
introduced for special solutions related to topological conformal
field theories \cite{bib:Krichever,bib:Dubrovin}, then reformulated
for general solutions in the following form \cite{bib:TT-dKP}.

In the language of the dispersionless KP hierarchy, the F function
$F = F(t,x)$ can be characterized as a solution of the equations
\beqn
    \dfrac{\rd F}{\rd t_n} = h_n, \quad
    \dfrac{\rd F}{\rd x} = h_1,
                                           \label{eq:dKPDefEqF1}
\eeqn
or, equivalently, as a (local) potential of a closed 1-form,
\beqn
    dF = \sum_{n=1}^\infty h_n dt_n + h_1 dx.
                                           \label{eq:dKPDefEqF2}
\eeqn
Frobenius integrability of (\ref{eq:dKPDefEqF1}), or closedness
of the right hand side of (\ref{eq:dKPDefEqF2}), is indeed ensured
by the dispersionless KP hierarchy itself. Now $h_n$'s are derivatives
of a single function, $F$. Meanwhile, it is also known that all $g_n$'s
are differential polynomials of $h_n$'s; $g_n$'s, too, can be reproduced
from $F$. Thus $F$ plays the role of a ``generating function''. This
role of $F$ is parallel to the tau function of the full KP hierarchy.

The $F$ function has another characterization. The full KP hierarchy
can be reformulated so as to include a Planck constant $\hbar$.
The tau function, too, then depends on $\hbar$ as $\tau =
\tau(\hbar,t,x)$. In the limit of $\hbar \to 0$, $\tau$ behaves as
\beqn
    \tau(\hbar,t,x)
    = \exp\left( - \hbar^{-2} F(t,x) + O(\hbar^{-1}) \right),
                                        \label{eq:KPAsympTau}
\eeqn
thus $F$ emerges as the leading term of $\hbar$-expansion of
$\log\tau$.

In the context of differential rings, defining $F$ as above
amounts to extending the differential rings $\cR_0$ and $\cR$
into the rings $\cR_0[F]$ and $\cR[F]$ of polynomials in $F$
(with coefficients taken from $\cR_0$ and $\cR$). One should
then reinterpret Eqs. (\ref{eq:dKPDefEqF1}) and
(\ref{eq:dKPDefEqF2}) as defining an extension of the abstract
derivations $\rd$ ($\leftrightarrow \rd/\rd x$) and
$\rd_n$ ($\leftrightarrow \rd/\rd t_n$) of $\cR$ onto these
larger rings; Frobenius integrability condition (or closedness
of the 1-form) is nothing more than a consistency condition for
this extension of derivations.

This point of view also applies to symmetries. Indeed the symmetries
$\delta_\cA$ on $\cR$, too, turn out to have a consistent extension
onto $\cR$ as follows:
\beqn
    \delta_A F
    = - \res_\lambda \int_0^{\cM(\lambda)} d\mu \cA(\lambda,\mu),
                                                \label{eq:dKPSymmF}
\eeqn
where ``$\res_\lambda$'' means the formal residue with respect to
$\lambda$:
\beqn
    \res_\lambda \sum a_n \lambda^n = a_{-1}.
\eeqn
Note that the above formula can also be written as a double integral
over a domain of a two-dimensional ``phase space'' (a cylinder) with
coordinates $(\lambda,\mu)$:
\beqn
    \delta_A F
    =  - \oint \frac{d\lambda}{2\pi i}
         \int_0^{\cM(\lambda)} d\mu \cA(\lambda,\mu),
                                             \label{eq:dKPSymmFbis}
\eeqn
where the path of the $\lambda$-integral is understood to be a small
loop around $\lambda = \infty$. It is interesting that this resembles
a ``fermi fluid'' picture of $c=1$ matrix models \cite{bib:fermi-fluid}.
We shall discuss this issue in more detail in the final section.

Consistency conditions of the above extension of $\delta_\cA$
are given by the equations
\beqnarray
    \dfrac{\rd}{\rd t_n} \delta_\cA F
      &=& \delta_\cA \dfrac{\rd F}{\rd t_n}
         \quad (=  \delta_\cA h_n),
                                              \nonumber \\
    \dfrac{\rd}{\rd x} \delta_\cA F
      &=& \delta_\cA \dfrac{\rd F}{\rd x}
         \quad (=  \delta_\cA h_1),
\eeqnarray
which simply say that $\delta_\cA$ commutes with $\rd/\rd t_n$
and $\rd/\rd x$ as derivations on $\cR[F]$.  One can indeed prove
that (\ref{eq:dKPSymmF}) satisfies these conditions.

The extended symmetries obey the ``anomalous'' commutation relations
\beqn
    [\delta_\cA, \delta_\cB ]
    = \delta_{ \{ \cA,\cB \}_{\lambda\mu} } + c(\cA,\cB)\rd_F,
                                          \label{eq:dKPextCommRel}
\eeqn
where $c(\cA,\cB)$ is a 2-cocycle of the two-dimensional Poisson
algebra,
\beqn
    c(\cA,\cB)
    = \res_\lambda A(\lambda,0) \frac{\rd \cB(\lambda,0)}{\rd\lambda},
\eeqn
and $\rd_F$ is yet another abstract derivation on $\cR$ defined by
\beqn
    \rd_F(F) = 1, \quad
    \rd_F (g_n^{(\alpha)}) = \rd_F (h_n^{(\alpha)})
    = \rd_F (t_n) = \rd_F (x) = 0.
\eeqn
Actually, (\ref{eq:dKPextCommRel}) is nothing but (\ref{eq:dKPCommRel})
plus the following relation:
\beqn
    [\delta_\cA, \delta_\cB ] F
    = \delta_{ \{ \cA,\cB \}_{\lambda\mu} } F + c(\cA,\cB).
\eeqn
Note that $\rd_F$ commutes with all other derivations $\rd/\rd t_n$,
$\rd/\rd x$ and $\delta_\cA$. Thus the extended symmetries obey a
central extension of the two-dimensional Poisson algebra.

\section{Toroidal model of higher dimensional dispersionless
integrable hierarchies}

\subsection{Formulation of toroidal model}

We now introduce the toroidal model of higher dimensional
dispersionless integrable hierarchies. The construction is
very similar to the planar model in our previous work
\cite{bib:T-MoyalKP}.

Let $\theta = (\theta_1,\theta_2)$ be angle variables of a
two-dimensional torus $T^2$ ($\theta_j \sim \theta_j + 2 \pi n$,
$n=\mbox{integer}$, $j = 1,2$).  Any function $A = A(\theta)$ on
$T^2$ can be expanded into a Fourier series ($i = \sqrt{-1}$),
\beqn
    A = \sum_{\alpha_1,\alpha_2=-\infty}^\infty
        a_{\alpha_1 \alpha_2}
        e^{i(\alpha_1 \theta_1 + \alpha_2\theta_2)}.
\eeqn
For such functions, we define the Poisson bracket
\beqn
    \{ A, B \}_\theta
    =  \dfrac{\rd A}{\rd \theta_1} \dfrac{\rd B}{\rd \theta_2}
     - \dfrac{\rd A}{\rd \theta_2} \dfrac{\rd B}{\rd \theta_1}.
\eeqn
Furthermore, if $A$ and $B$ also depend on two other variables
$(k,x)$ (which are ``phase space variables'' of the
dispersionless KP hierarchy), we define the four-dimensional
Poisson bracket
\beqn
    \{ A, B \} = \{ A, B \}_{kx} + \{ A, B \}_\theta.
\eeqn

The toroidal model is defined by the Lax equations
\beqnarray
  \dfrac{\rd \cL}{\rd t_{n\alpha}} = \{ \cB_{n\alpha}, \cL \}, &\quad&
  \dfrac{\rd \cM}{\rd t_{n\alpha}} = \{ \cB_{n\alpha}, \cM \},
                                                  \nonumber \\
  \dfrac{\rd \cU}{\rd t_{n\alpha}} = \{ \cB_{n\alpha}, \cU \}, &\quad&
  \dfrac{\rd \cV}{\rd t_{n\alpha}} = \{ \cB_{n\alpha}, \cV \},
                                                  \label{eq:Lax}
\eeqnarray
and the canonical relations
\beqnarray
  \{ \cL, \cM \} = \{ \cU, \cV \} &=& 1,          \nonumber \\
  \{ \cL, \cU \} = \{ \cL, \cV \} &=&
  \{ \cM, \cU \} = \{ \cM, \cV \} = 0.
                                                  \label{eq:CanRel}
\eeqnarray
The time variables $t = (t_{n\alpha})$ have a double index
$(n,\alpha)$  with $n = 0,1,\ldots$ and
$\alpha = 0, \pm 1, \pm 2, \ldots$.  In addition to the analogues
$\cL$ and $\cM$ of the Lax-Orlov-Shulman functions in the
dispersionless KP hierarchy, we now have two extra functions $\cU$
and $\cV$ that correspond to the two extra spatial dimensions.
These four functions are assumed to be Laurent-Fourier series
of the form
\beqnarray
  \cL &=& k + \sum_{n=1}^\infty g_{n+1}(t,x,\theta) k^{-n},
                                                 \nonumber \\
  \cM &=& \sum_{n,\alpha} n t_{n\alpha} \cL^{n-1} e^{i\alpha\cU}
          + x
          + \sum_{n=1}^\infty h_n(t,x,\theta) \cL^{-n-1},
                                                 \nonumber \\
  \cU &=& \theta_1 + \sum_{n=1}^\infty u_n(t,x,\theta) \cL^{-n},
                                                 \nonumber \\
  \cV &=& \sum_{n,\alpha} i \alpha t_{n\alpha} \cL^n e^{i\alpha\cU}
          + \theta_2
          + \sum_{n=1}^\infty v_n(t,x,\theta) \cL^{-n},
\eeqnarray
where the coefficients $g_n$, $h_n$, $u_n$ and $v_n$ are required
to be functions on $T^2$ (i.e., doubly periodic). Note that $\cU$
and $\cV$ themselves are not doubly periodic (because of the presence
of the $\theta_1$ and $\theta_2$ terms on the right hand side), but
their exponentials $e^{i\alpha\cU}$ are doubly periodic. Finally,
$\cB_{n\alpha}$ are given by
\beqn
    \cB_{n\alpha} = \left( \cL^n e^{i\alpha\cU} \right)_{\ge 0},
\eeqn
where $(\quad)_{\ge 0}$ and $(\quad)_{\le -1}$ are the same as those
of the dispersionless KP hierarchy, i.e., the projections onto
nonnegative and negative powers of $k$.

A few technical remarks are now in order:

1. The construction of the toroidal model is almost parallel to the
planar model. A main difference is that whereas the planar model
is based on Tayler series, the toroidal model uses Fourier series.
In the planar model, the angle variables $(\theta_1,\theta_2)$
are replaced by planar coordinates $(y,z)$, and the counterparts
of $\cL$, $\cM$, $\cU$ and $\cV$ are assumed to have the following
expansion:
\beqnarray
  \cL &=& k + \sum_{n=1}^\infty g_{n+1}(t,x,y,z) k^{-n},
                                                 \nonumber \\
  \cM &=& \sum_{n,\alpha} n t_{n\alpha} \cL^{n-1} \cU^\alpha
          + x
          + \sum_{n=1}^\infty h_n(t,x,y,z) \cL^{-n-1},
                                                 \nonumber \\
  \cU &=& y + \sum_{n=1}^\infty u_n(t,x,y,z) \cL^{-n},
                                                 \nonumber \\
  \cV &=& \sum_{n,\alpha} \alpha t_{n\alpha} \cL^n \cU^{\alpha-1}
          + z
          + \sum_{n=1}^\infty v_n(t,x,y,z) \cL^{-n}.
\eeqnarray
Furthermore, the second index $\alpha$ is restricted to nonnegative
values $\alpha = 0, 1, \ldots$, and $\cB_{n\alpha}$ are given by
\beqn
    \cB_{n\alpha} = \left( \cL^n \cU^\alpha \right)_{\ge 0}.
\eeqn
Apart from this difference, both hierarchies can be treated in
much the same way.

2. It is easy to see that the above extended Lax system indeed
gives a well defined set of flows in the space of four functions
$(\cL,\cM,\cU,\cV)$.
First, by the definition of $\cB_{n\alpha}$ and the canonical
relations among $(\cL,\cM,\cU,\cV)$, the Lax equations of $\cL$
and $\cU$ can be rewritten
\beqnarray
  \dfrac{\rd \cL}{\rd t_{n\alpha}}
    &=& - \{ \cB_{n\alpha}^{-}, \cL \},
                                                  \nonumber \\
  \dfrac{\rd \cU}{\rd t_{n\alpha}}
    &=& - \{ \cB_{n\alpha}^{-}, \cU \},
\eeqnarray
where we have defined
\beqn
    \cB_{n\alpha}^{-} = \left(\cL^n e^{i\alpha\cU}\right)_{\le -1}.
\eeqn
Both hand sides of these equations contain only negative powers
of $k$, and one can read off a closed set of equations of motions
for $g_n$ and $u_n$ therefrom.  Second, as for $\cM$ and $\cV$,
let us drop the first part containing $t$'s and consider
\beqnarray
    \dfrac{\rd \cM'}{\rd t_{n\alpha}}
      &=& \{ \cB_{n\alpha}, \cM' \} - b \cL^{n-1} e^{i\alpha\cU}
       = \{ \cB_{n\alpha}^{-}, \cM' \},
                                                 \nonumber \\
    \dfrac{\rd \cV'}{\rd t_{n\alpha}}
      &=& \{ \cB_{n\alpha}, \cV' \} - i \alpha \cL^n e^{i\alpha\cU}
       = \{ \cB_{n\alpha}^{-}, \cV' \},
\eeqnarray
These equations, too, contain only negative powers of $k$, and
give a set of equations of motions for $h_n$ and $v_n$. Third,
in terms of these modified quantities, canonical relations
(\ref{eq:CanRel}) become
\beqnarray
  \{ \cL, \cM' \} = \{ \cU, \cV' \} &=& 1,          \nonumber \\
  \{ \cL, \cU \} = \{ \cL, \cV' \}  &=&
  \{ \cM', \cU \} = \{ \cM', \cV' \} = 0.
\eeqnarray
These relations have to be imposed on the above equations of motion
for $(\cL,\cM',\cU,\cV')$.  Since the above equations of motion take
the form of ``Liouville flows'' of classical mechanics, the last
constraints are indeed consistent with the flows (i.e., preserved
for all time if satisfied at the initial time $t = 0$).

3. One can derive the zero-curvature equations
\beqn
    \dfrac{\rd \cB_{m\alpha}}{\rd t_{n\beta}}
    - \dfrac{\rd \cB_{n\beta}}{\rd t_{m\alpha}}
    + \{ \cB_{m\alpha}, \cB_{n\alpha} \} = 0
                                             \label{eq:ZeroCurv}
\eeqn
from the Lax equations of $\cL$ and $\cU$.  This can be seen by the
same method as used in the full and dispersionless KP hierarchy
as follows:  Note that any monomial of $\cL$ and $e^{i\cU}$ obeys
the same Lax equations:
\beqn
    \dfrac{\rd}{\rd t_{n\beta}} (\cL^m e^{i\alpha \cU})
    = \{ \cB_{n\beta}, \cL^m e^{i\alpha \cU} \}.
\eeqn
Subtracting from this the same equation with $(m,\alpha)$ and
$(n,\beta)$ interchanged, one has
\beqn
      \dfrac{\rd}{\rd t_{n\beta}} (\cL^m e^{i\alpha \cU})
    - \dfrac{\rd}{\rd t_{m\alpha}} (\cL^n e^{i\beta \cU})
    = \{ \cB_{n\beta}, \cL^m e^{i\alpha \cU} \}
    - \{ \cB_{m\alpha}, \cL^n e^{i\beta \cU} \}.
                                     \label{eq:ZeroCurvProof}
\eeqn
Let us consider both hand sides in more detail.
Since $\cL^m e^{i\alpha\cU} = \cB_{m\alpha} + \cB_{m\alpha}^{-}$,
LHS becomes a sum of two parts that consist of only nonnegative
or negative powers of $k$:
\beqn
    \mbox{LHS of (\ref{eq:ZeroCurvProof})}
    = \left( \dfrac{\rd \cB_{m\alpha}}{\rd t_{n\beta}}
           - \dfrac{\rd \cB_{n\beta}}{\rd t_{m\alpha}} \right)
    + \left( \dfrac{\rd \cB_{m\alpha}^{-}}{\rd t_{n\beta}}
           - \dfrac{\rd \cB_{n\beta}^{-}}{\rd t_{m\alpha}} \right).
\eeqn
Similarly, recalling the Poisson commutativity of $\cL$ and $\cU$
again, one can rewrite the right hand side as:
\beqnarray
    \mbox{RHS of (\ref{eq:ZeroCurvProof})}
    &=&   \{ \cB_{n\beta}, \cL^m e^{i\alpha \cU} \}
        + \{ \cB_{m\alpha}^{-}, \cL^n e^{i\beta \cU} \}
                                                       \nonumber \\
    &=&   \{ \cB_{n\beta}, \cB_{m\alpha} \}
        + \{ \cB_{n\beta}, \cB_{m\alpha}^{-} \}
        + \{ \cB_{m\alpha}^{-}, \cB_{n\beta} \}
        + \{ \cB_{m\alpha}^{-}, \cB_{n\beta}^{-} \}
                                                       \nonumber \\
    &=&   \{ \cB_{n\beta}, \cB_{m\alpha} \}
        + \{ \cB_{m\alpha}^{-}, \cB_{n\beta}^{-} \}.
\eeqnarray
Thus both hand sides become a sum of two parts each of which contains
only nonnegative or negative powers of $k$. The $(\quad)_{\ge 0}$
part gives zero-curvature equation (\ref{eq:ZeroCurv}),
whereas the $(\quad)_{\le -1}$ part yields its ``dual'' form,
\beqn
    \dfrac{\rd \cB_{m\alpha}^{-}}{\rd t_{n\beta}}
    - \dfrac{\rd \cB_{n\beta}^{-}}{\rd t_{m\alpha}}
    - \{ \cB_{m\alpha}^{-}, \cB_{n\alpha}^{-} \} = 0.
\eeqn

\subsection{Integrability, twistor theory and Riemann-Hilbert problem}

Just as in the case of the planar model, integrability of this new
hierarchy can be explained in two different ways.

One way is to construct a new hierarchy in which the Poisson bracket
is replaced by a Moyal bracket, and to resort to integrability
of the latter. The original model can be reproduced in the classical
limit $\hbar \to 0$. In our previous paper \cite{bib:T-MoyalKP},
the planar model was treated exactly along this line of approach.

Another way is to search for a twistor theoretical representation (or
an equivalent Riemann-Hilbert problem) of the system.  This method
relies on the existence of a 2-form equation like Eq.
(\ref{eq:dKP2-form}) of the dispersionless KP hierarchy.
One can indeed find such a 2-form equation as follows.

\begin{proposition} \label{prop:Lax<=>2-form}
The extended Lax system of the toroidal model is equivalent to the
2-form equation
\beqn
    d\cL \wedge d\cM + d\cU \wedge d\cV
    = dk \wedge dx + d\theta_1 \wedge d\theta_2
      + \sum_{n,\alpha} d\cB_{n\alpha} \wedge dt_{n\alpha}.
                                           \label{eq:2-form}
\eeqn
\end{proposition}

A proof is given in the next subsection. We here briefly show
why this leads to integrability of the model. Basic ieas are
the same as in the case of the self-dual Einstein equation
\cite{bib:Boyer-Plebanski,bib:T-SDG} and subsequent applications
to the dispersionless KP and Toda hierarchies
\cite{bib:TT-dKP,bib:TT-dToda}, the toroidal model of higher
dimensional dispersionless hierarchy \cite{bib:T-MoyalKP}, etc.

Let $\omega$ denote the right hand side of (\ref{eq:2-form}).
Obviously $\omega$ is a closed form on the $(t,k,x,\theta)$ space,
\beqn
    d\omega = 0.
\eeqn
Furthermore, although the third order exterior power does not
vanish, $\omega \wedge \omega \wedge \omega \not= 0$,
the fourth one vanishes:
\beqn
    \omega \wedge \omega \wedge \omega \wedge \omega = 0.
\eeqn
(One can prove this using zero-curvature equations
(\ref{eq:ZeroCurv}).) Therefore, by a theorem of Darboux in
symplectic geometry, $\omega$ is a degenerate symplectic form
and has a local expression
\beqn
    \omega = d\cP_1 \wedge d\cQ_1 + d\cP_2 \wedge d\cQ_2
\eeqn
with suitable four functions $(\cP_1,\cQ_1,\cP_2,\cQ_2)$
(Darboux variables). The above proposition shows that
$(\cL,\cM,\cU,\cV)$ give such a Darboux variable system, which
is valid in a neighborhood of $k = \infty$ (i.e., gives a set
of complex analytic functions defined in a neighborhood of
$k = \infty$ except at $k = \infty$ itself). In fact, the theorem
of Darboux ensures only local existence, and one can find another
Darboux variable system, say $(\cLbar,\cMbar,\cUbar,\cVbar)$,
that are similarly valid in a neighborhood of $k = 0$.

Suppose now that the domains $D$ and $\Dbar$ of these two Darboux
variable systems in the $k$ plane (or Riemann sphere) are large
enough to cover the whole Riemann sphere. (This is indeed a generic
situation of Penrose's nonlinear graviton construction.) The two
Darboux variable systems are then linked, via $\omega$, as
\beqn
    d\cL \wedge d\cM + d\cU \wedge d\cV
    = \omega
    = d\cLbar \wedge d\cMbar + d\cUbar \wedge d\cVbar.
\eeqn
This implies that they are functionally related,
\beqnarray
    \cLbar = f_1(\cL,\cM,\cU,\cV), &&
    \cMbar = f_2(\cL,\cM,\cU,\cV),      \nonumber \\
    \cUbar = f_3(\cL,\cM,\cU,\cV), &&
    \cVbar = f_4(\cL,\cM,\cU,\cV),
                                     \label{eq:RHproblem}
\eeqnarray
and that the quartet $f = (f_1,f_2,f_3,f_4)$ becomes a
four-dimensional symplectic diffeomorphism
$(\lambda,\mu,u,v) \mapsto f(\lambda,\mu,u,v)$, i.e.,
satisfies the symplectic condition
\beqn
    df_1 \wedge df_2 + df_3 \wedge df_4
    = d\lambda \wedge d\mu + du \wedge dv.
\eeqn
In the geometric language of twistor theory \cite{bib:twistor},
this symplectic diffeomorphism may be interpreted as a transition
function between two local coordinate patches of a four-dimensional
complex manifold (twistor space); Eq. (\ref{eq:RHproblem}) then
define a family of holomorphic curves in the twistor space.
{}From the point of view of the theory of integrable systems,
Eq. (\ref{eq:RHproblem}) may be thought of as a kind of nonlinear
Riemann-Hilbert problem.

This Riemann-Hilbert problem, like those for the other equations
mentioned above, is ensured to have a unique solution as far as
$f$ is sufficiently close to the identity diffeomorphism.
(Actually, unique solvability is also preserved under small
perturbations from any soluble data. In the language of twistor
theory, this is a consequence of the Kodaira-Spencer theory
\cite{bib:twistor}.)  This means that there is a one-to-one
correspondence between a solution and its twistor data. Such
a correspondence lies in the heart of integrable nature of
almost all integrable systems. In this sense, the new hierarchy,
too, may be called ``integrable''.

\subsection{Proof of Proposition}

We give a proof of Proposition \ref{prop:Lax<=>2-form}.
The proof is somewhat lengthy and technical, but basic ideas
are mostly similar to the case of the dispersionless KP hierarchy.

Let us first derive the extended Lax system from the 2-form equation.
Note that the total differentials $d\cL$ etc. can be expanded as
\beqn
    d\cL = \frac{\rd \cL}{\rd k} dk
         + \frac{\rd \cL}{\rd x} dx
         + \dfrac{\rd \cL}{\rd \theta_1} d\theta_1
         + \dfrac{\rd \cL}{\rd \theta_1} d\theta_2
         + \sum_{n,\alpha}
               \dfrac{\rd \cL}{\rd t_{n\alpha}} dt_{n\alpha},
    \ldots .
\eeqn
Substituting this expression, one can write the the left hand side
of the 2-form equation as a linear combination of the basic 2-forms
$dk \wedge dx$, etc.  We now pick out coefficients for each basic
2-form.

First, modulo terms including $dt$'s, the 2-form equation
can be written
\beqn
    d\cL \wedge d\cM + d\cU \wedge d\cV
      = dk \wedge dx + d\theta_1 \wedge d\theta_2
        + (\mbox{terms including $dt$'s}).
\eeqn
This shows that the four-dimensional map
$(k,x,\theta_1,\theta_2) \mapsto (\cL,\cM,\cU,\cV)$ is a
symplectic diffeomorphism with parameters $t_{n\alpha}$.
Therefore canonical relations (\ref{eq:CanRel}) follow.

Second, from terms of the form $dk \wedge dt_{n\alpha}$,
$dx \wedge dt_{n\alpha}$ and $d\theta_j \wedge dt_{n\alpha}$ ,
respectively, the following equations emerge:
\beqnarray
    \dfrac{\rd(\cL,\cM)}{\rd(k,t_{n\alpha})}
      + \dfrac{\rd(\cU,\cV)}{\rd(k,t_{n\alpha})}
         &=& \dfrac{\rd \cB_{n\alpha}}{\rd k},
                                            \nonumber \\
    \dfrac{\rd(\cL,\cM)}{\rd(x,t_{n\alpha})}
      + \dfrac{\rd(\cU,\cV)}{\rd(x,t_{n\alpha})}
         &=& \dfrac{\rd \cB_{n\alpha}}{\rd x},
                                            \nonumber \\
    \dfrac{\rd(\cL,\cM)}{\rd(\theta_j,t_{n\alpha})}
      + \dfrac{\rd(\cU,\cV)}{\rd(\theta_j,t_{n\alpha})}
         &=& \dfrac{\rd \cB_{n\alpha}}{\rd \theta_j},
\eeqnarray
where we have used the notation
\beqn
    \frac{\rd(A,B)}{\rd(p,q)}
      = \frac{\rd A}{\rd p} \frac{\rd B}{\rd q}
        - \frac{\rd A}{\rd p} \frac{\rd B}{\rd q}.
\eeqn
These four equations give a linear system of equations for the
$t$-derivatives of $\cL$, $\cM$, $\cU$ and $\cV$:
\beqn
     \left( \dfrac{\rd \cM}{\rd t_{n\alpha}},
              - \dfrac{\rd \cL}{\rd t_{n\alpha}},
                  \dfrac{\rd \cV}{\rd t_{n\alpha}},
                    - \dfrac{\rd \cU}{\rd t_{n\alpha}} \right)
                      J
   =  \left( \dfrac{\rd \cB_{n\alpha}}{\rd k},
               \dfrac{\rd \cB_{n\alpha}}{\rd x},
                 \dfrac{\rd \cB_{n\alpha}}{\rd \theta_1},
                   \dfrac{\rd \cB_{n\alpha}}{\rd \theta_2} \right),
 \eeqn
where the matrix $J$ is given by
\beqn
    J = \left(
          \begin{array}{rrrr}
            \dfrac{\rd \cL}{\rd k} &
              \dfrac{\rd \cL}{\rd x} &
                \dfrac{\rd \cL}{\rd \theta_1} &
                  \dfrac{\rd \cL}{\rd \theta_2} \\[2mm]
            \dfrac{\rd \cM}{\rd k} &
              \dfrac{\rd \cM}{\rd x} &
                \dfrac{\rd \cM}{\rd \theta_1} &
                  \dfrac{\rd \cM}{\rd \theta_2} \\[2mm]
            \dfrac{\rd \cU}{\rd k} &
              \dfrac{\rd \cU}{\rd x} &
                \dfrac{\rd \cU}{\rd \theta_1} &
                  \dfrac{\rd \cU}{\rd \theta_2} \\[2mm]
            \dfrac{\rd \cV}{\rd k} &
              \dfrac{\rd \cV}{\rd x} &
                \dfrac{\rd \cV}{\rd \theta_1} &
                  \dfrac{\rd \cV}{\rd \theta_2}
          \end{array}
        \right).
\eeqn
Since the four-dimensional map $(k,x,\theta_1,\theta_2)
\mapsto (\cL,\cM,\cU,\cV)$ is symplectic, this matrix $J$
is a symplectic matrix, and the inverse matrix can be written
\beqn
    J^{-1}
      = \left(
          \begin{array}{rrrr}
            \dfrac{\rd \cM}{\rd x} &
              - \dfrac{\rd \cL}{\rd x} &
                  \dfrac{\rd \cV}{\rd x} &
                    - \dfrac{\rd \cU}{\rd x} \\[2mm]
            - \dfrac{\rd \cM}{\rd k} &
                \dfrac{\rd \cL}{\rd k} &
                  - \dfrac{\rd \cV}{\rd k} &
                      \dfrac{\rd \cU}{\rd k} \\[2mm]
            \dfrac{\rd \cM}{\rd \theta_2} &
              - \dfrac{\rd \cL}{\rd \theta_2} &
                  \dfrac{\rd \cV}{\rd \theta_2} &
                     - \dfrac{\rd \cU}{\rd \theta_2} \\[2mm]
            - \dfrac{\rd \cM}{\rd \theta_1} &
                \dfrac{\rd \cL}{\rd \theta_1} &
                  - \dfrac{\rd \cV}{\rd \theta_1} &
                      \dfrac{\rd \cU}{\rd \theta_1}
          \end{array}
        \right).
\eeqn
Therefore the above equations can be solved for $\cL$, $\cM$,
$\cU$ and $\cV$ as follows:
\beqn
    \left( \dfrac{\rd \cM}{\rd t_{n\alpha}},
              - \dfrac{\rd \cL}{\rd t_{n\alpha}},
                  \dfrac{\rd \cV}{\rd t_{n\alpha}},
                    - \dfrac{\rd \cU}{\rd t_{n\alpha}} \right)
   = \left( \dfrac{\rd \cB_{n\alpha}}{\rd k},
              \dfrac{\rd \cB_{n\alpha}}{\rd x},
                \dfrac{\rd \cB_{n\alpha}}{\rd \theta_1},
                  \dfrac{\rd \cB_{n\alpha}}{\rd \theta_2}
                     \right) J^{-1}.
\eeqn
Plugging the previous expression of $J^{-1}$ and calculating
the right hand side explicitly, one will find that this gives
Lax equations (\ref{eq:Lax}) exactly.

Finally, terms of the form $dt_{n\beta} \wedge dt_{m\alpha}$ give
\beqn
      \dfrac{\rd(\cL,\cM)}{\rd(t_{m\alpha},t_{n\beta})}
    + \dfrac{\rd(\cU,\cV)}{\rd(t_{m\alpha},t_{n\beta})}
  =   \dfrac{\rd \cB_{n\beta}}{\rd t_{m\alpha}}
    - \dfrac{\rd \cB_{m\alpha}}{\rd t_{n\beta}}.
\eeqn
Using (\ref{eq:Lax},\ref{eq:CanRel}), which we have proven, one
can easily see that the left hand side coincides with
$\{ \cB_{m\alpha}, \cB_{n\beta} \}$. The above equation therefore
turns out to be zero-curvature equation (\ref{eq:ZeroCurv}).
Thus all equations of the extended Lax system can be derived
from the 2-form equation.

One can easily see that the converse is also true. Recall that
zero-curvature equations (\ref{eq:ZeroCurv}) can be derived from
the Lax equations. This ensures the vanishing of $dt_{n\beta}
\wedge dt_{m\beta}$-part in the 2-form equation. Having this,
one can actually check that the above derivation is reversible.

\subsection{Another expression of hierarchy}

Both the planar and toroidal models are formulated in terms
of the four fundamental quantities $(\cL,\cM,\cU,\cV)$. This
expression is particularly suited for understanding the twistor
theoretical meaning and the underlying Riemann-Hilbert problem.
In the subsequent sections, however, another expression of the
same hierarchy becomes more useful.

Let us introduce a new variable $\lambda$ (``spectral parameter'')
and define the following Laurent series of $\lambda$:
\beqnarray
    \cM(\lambda)
      &=& \sum_{n,\alpha}
             n t_{n\alpha} \lambda^{n-1} e^{i\alpha \cU(\lambda)}
        + x + \sum_{n=1}^\infty h_n \lambda^{-n-1},
                                                   \nonumber \\
    \cU(\lambda)
      &=& \theta_1 + \sum_{n=1}^\infty u_n \lambda^{-n},
                                                   \nonumber \\
    \cV(\lambda)
      &=& \sum_{n,\alpha}
             i \alpha t_{n\alpha} \lambda^n e^{i\alpha \cU(\lambda)}
        + \theta_2 + \sum_{n=1}^\infty v_n \lambda^{-n},
                                                   \nonumber \\
    \cB_{n\alpha}(\lambda)
      &=& \lambda^n e^{i\alpha \cU(\lambda)}
        + \sum_{n=1}^\infty f_{n\alpha,m+1} \lambda^{-m},
\eeqnarray
where $g_n$, $u_n$ and $v_n$ are the same as in the definition
of $\cM$, $\cU$ and $\cV$. Siminarly, the coefficients of
$\cB_{n\alpha}(\lambda)$ are those of Laurent expansion of
$\cB_{n\alpha}$ in $\cL$:
\beqn
    \cB_{n\alpha}
    = \cL^n e^{i\alpha \cU(\lambda) }
      + \sum_{n=1}^\infty f_{n\alpha,m+1} \cL^{-m}.
\eeqn

To see that $\cB_{n\alpha}$ has such Laurent expansion,
let us recall that $\cL$ is a Laurent series of $k$ of the form
\beqn
    \cL = k + \sum_{n=1}^\infty g_{n+1}(t,x,\theta) k^{-n}.
\eeqn
This relation can be solved for $k$,
\beqn
    k = \cL + \sum_{n=1}^\infty f_{n+1}(t,x,\theta) \cL^{-n}.
\eeqn
(In the case of the dispersionless KP hierarchy, these coefficients
$f_n$ are directly related with ``flat coordinates'' of topological
Landau-Ginzburg models \cite{bib:Krichever,bib:Dubrovin,bib:TT-dKP}.)
Since $\cB_{n\alpha}$ is a polynomial in $k$, substitution of this
expression of $k$ yields an expansion of $\cB_{n\alpha}$ as
mentioned above. Replacing $\cL \to \lambda$, one obtains
$\cB_{n\alpha}(\lambda)$.  In particular, since $\cB_{10} = k$,
$\cB_{10}(\lambda)$ turns out to be written
\beqn
    \cB_{10}(\lambda)
      = \lambda + \sum_{n=1}^\infty f_{n+1} \lambda^{-n}.
\eeqn

Conversely, one can reproduce $\cL$, $\cM$, $\cU$ and $\cV$ from the
above Laurent series of $\lambda$ by solving the relation
\beqn
    k = \cB_{10}(\lambda)
\eeqn
for $\lambda$. this gives $\lambda = \cL$ as a function of
$(t,k,x,\theta)$. Substituting $\lambda = \cL$ in $\cU(\lambda)$,
$\cU(\lambda)$ and $\cV(\lambda)$ then give $\cM$, $\cU$ and $\cV$.

The extended Lax system, too, can be converted into the language
of the Latter Laurent series:

\begin{proposition} \label{prop:Lax<=>Lax(lambda)}
Extended Lax system (\ref{eq:Lax},\ref{eq:CanRel}) is equivalent to
the following system consisting of three different sets of equations.
\newline
a) Lax equations
\beqnarray
  && \dfrac{\rd \cU(\lambda)}{\rd t_{n\alpha}}
       = \{ \cB_{n\alpha}(\lambda), \cU(\lambda) \}_\theta,
                                                  \nonumber \\
  && \dfrac{\rd \cV(\lambda)}{\rd t_{n\alpha}}
       = \{ \cB_{n\alpha}(\lambda), \cV(\lambda) \}_\theta,
                                                  \nonumber \\
  && \dfrac{\rd \cM(\lambda)}{\rd t_{n\alpha}}
       = \dfrac{\rd \cB_{n\alpha}(\lambda)}{\rd \lambda}
            + \{ \cB_{n\alpha}(\lambda), \cM(\lambda) \}_\theta,
                                     \label{eq:Lax(lambda)}
\eeqnarray
b) canonical relations (and related equations)
\beqnarray
  && \{ \cU(\lambda), \cV(\lambda) \}_\theta
       = 1,
                                                  \nonumber \\
  && \{ \cM(\lambda), \cU(\lambda) \}_\theta
       = \dfrac{\rd \cU(\lambda)}{\rd \lambda},
                                                  \nonumber \\
  && \{ \cM(\lambda), \cV(\lambda) \}_\theta
       = \dfrac{\rd \cV(\lambda)}{\rd \lambda},
                                  \label{eq:CanRel(lambda)}
\eeqnarray
c) zero-curvature equations
\beqn
     \dfrac{\rd \cB_{m\alpha}(\lambda)}{\rd t_{n\beta}}
       - \dfrac{\rd \cB_{n\beta}(\lambda)}{\rd t_{m\alpha}}
         + \{ \cB_{m\alpha}(\lambda), \cB_{n\beta}(\lambda) \}_\theta
           = 0.
                                  \label{eq:ZeroCurv(lambda)}
\eeqn
Here (\ref{eq:Lax(lambda)}) is understood to include differential
equations with respect to $x$, too, which are identical to the
differential equations with respect to $t_{10}$.
\end{proposition}

\begin{remark}
Basically the same is also true for the planar model (except that
the toroidal Poisson bracket $\{\quad,\quad\}_\theta$ has to be
replaced by the planar Poisson bracket $\{\quad,\quad\}_{yz}$).
The dispersionless KP hierarchy, too, has a similar expression:
\beqnarray
    \dfrac{\rd \cM(\lambda)}{\rd t_n}
       &=& \dfrac{\rd \cB_n(\lambda)}{\rd \lambda},
                                          \nonumber \\
    \dfrac{\rd \cB_m(\lambda)}{\rd t_n}
       &=& \dfrac{\rd \cB_n(\lambda)}{\rd t_m}.
\eeqnarray
Note that this is a condition for the existence of a potential.
Actually, the potential is given by the $S$ function $S(\lambda)$
\cite{bib:Krichever,bib:Dubrovin} i.e., the ``phase function'' of
a quasi-classical (WKB) expression of the Baker-Akhiezer function
\cite{bib:Kodama-Gibbons,bib:TT-qcKP}. This is not the case of the
higher dimensional hierarchy (because of the presence of the
Poisson bracket terms in (\ref{eq:Lax(lambda)}), indicating a crucial
difference between those higher and lower dimensional integrable
hierarchies.
\end{remark}

\subsection{Proof of Proposition}

We here give a proof of Proposition \ref{prop:Lax<=>Lax(lambda)}.
Since extended Lax system (\ref{eq:Lax}) is equivalent to 2-form
equation ({\ref{eq:2-form}), it is sufficient to prove the
equivalence with the 2-form equation.  Let us derive
(\ref{eq:Lax(lambda)},\ref{eq:CanRel(lambda)},
\ref{eq:ZeroCurv(lambda)}) from the 2-form equation
(\ref{eq:2-form}).

We first make a change of coordinates $(t,k,x,\theta) \to
(t,\lambda,x,\theta)$ by
\beqn
    \lambda = \cL(k,x,\theta,t).
\eeqn
Then 2-form equation (\ref{eq:2-form}) turns into the following
form:
\beqn
    d\lambda \wedge d\cM(\lambda)
      + d\cU(\lambda) \wedge d\cV(\lambda)
        = d\cB_{10}(\lambda) \wedge dx
            + d\theta_1 \wedge d\theta_2
              + \sum_{n\alpha}
                  d\cB_{n\alpha}(\lambda) \wedge dt_{n\alpha}.
\eeqn
Both hand sides of this equation can be expanded into a linear
combination of basic 2-forms: $d\theta_1 \wedge d\theta_2$,
$d\lambda \wedge d\theta_j$, $d\theta_j \wedge dt_{n\alpha}$,
$d\lambda \wedge dt_{n\alpha}$, $dt_{m\alpha} \wedge dt_{n\beta}$,
and those including $dx$.  Relations among their coefficients will
then give partial differential equations for $\cM(\lambda)$,
$\cU(\lambda)$ and $\cV(\lambda)$.  Since $\cB_{10} = k$, $dx$ and
$dt_{10}$ arise only as a linear combination of the form of
$dx + dt_{10}$. Therefore differential equations including $\rd/\rd x$
should take the same form as those including $\rd/\rd t_{10}$;
we do not have to consider them independently.

1. From $d\theta_1 \wedge d\theta_2$,
\beqn
    \dfrac{\rd ( \cU(\lambda),\cV(\lambda) )}
    {\rd\left( \theta_1,\theta_2 \right)} = 1.
\eeqn
This is exactly the first equation of (\ref{eq:CanRel(lambda)}).

2. From $d\lambda \wedge d\theta_j$,
\beqn
      \dfrac{\rd \cM(\lambda)}{\rd \theta_j}
    + \dfrac{\rd ( \cU(\lambda),\cV(\lambda) )}
            {\rd ( \lambda,\theta_j )}
    = 0.
\eeqn
These equations can be rewritten into a matrix form:
\beqn
    \left( \dfrac{\rd \cV(\lambda)}{\rd \lambda},
         - \dfrac{\rd \cU(\lambda)}{\rd \lambda} \right)
    \left( \begin{array}{rr}
           \dfrac{\rd \cU(\lambda)}{\rd \theta_1} &
             \dfrac{\rd \cU(\lambda)}{\rd \theta_2} \\[2mm]
           \dfrac{\rd \cV(\lambda)}{\rd \theta_1} &
             \dfrac{\rd \cV(\lambda)}{\rd \theta_2}
    \end{array} \right)
  = \left( \dfrac{\rd \cM(\lambda)}{\rd \theta_1},
           \dfrac{\rd \cM(\lambda)}{\rd \theta_2} \right).
\eeqn
Since the determinant of the coefficient matrix coincides with
$\{ \cU(\lambda), \cV(\lambda) \}_\theta = 1$, these equations
can be explicitly solved for $\rd\cU(\lambda)/\rd\lambda$
and $\rd\cV(\lambda)/\rd\lambda$:
\beqn
    \left( \dfrac{\rd \cV(\lambda)}{\rd \lambda},
         - \dfrac{\rd \cU(\lambda)}{\rd \lambda} \right)
  = \left( \dfrac{\rd \cM(\lambda)}{\rd \theta_1},
           \dfrac{\rd \cM(\lambda)}{\rd \theta_2} \right)
    \left( \begin{array}{rr}
           \dfrac{\rd \cV(\lambda)}{\rd \theta_2} &
             - \dfrac{\rd \cU(\lambda)}{\rd \theta_2} \\[2mm]
           - \dfrac{\rd \cV(\lambda)}{\rd \theta_1} &
               \dfrac{\rd \cU(\lambda)}{\rd \theta_1}
    \end{array} \right).
\eeqn
They give the second and third equations of (\ref{eq:CanRel(lambda)}).

3. From $d\theta_j \wedge dt_{n\alpha}$,
\beqn
    \dfrac{\rd ( \cU(\lambda),\cV(\lambda) )}
          {\rd ( \theta_j, t_{n\alpha} )}
  = \dfrac{\rd \cB_{n\alpha}(\lambda)}{\rd \theta_j}.
\eeqn
These equation, too, can be solved for
$\rd \cU(\lambda)/\rd t_{n\alpha}$ and
$\rd \cV(\lambda)/\rd t_{n\alpha}$ in the same way.
This gives the first and second equation of (\ref{eq:Lax(lambda)}).

4. From $d\lambda \wedge dt_{n\alpha}$,
\beqn
    \dfrac{\rd \cM(\lambda)}{\rd t_{n\alpha}}
    + \dfrac{\rd ( \cU(\lambda), \cV(\lambda) )}
            {\rd(\lambda,t_{n\alpha})}
    = \dfrac{\rd \cB_{n\alpha}(\lambda)}{\rd \lambda}.
\eeqn
By using (\ref{eq:Lax(lambda)}), which have been proven above,
one can rewrite the second term on the left hand side as:
\beqnarray
    \dfrac{\rd ( \cU(\lambda), \cV(\lambda) )}
          {\rd ( \lambda,t_{n\alpha} )}
    &=& \{ \cM(\lambda), \cU(\lambda) \}_\theta
          \{ \cB_{n\alpha}(\lambda), \cV(\lambda) \}_\theta
                                                   \nonumber \\
    &&  - \{ \cM(\lambda), \cV(\lambda) \}_\theta
          \{ \cB_{n\alpha}(\lambda), \cV(\lambda) \}_\theta,
                                                   \nonumber
\eeqnarray
and by (\ref{eq:CanRel(lambda)}), furthermore,
\beqn
    = \{ \cM(\lambda), \cB_{n\alpha}(\lambda) \}_\theta.
\eeqn
Thus the third equation of (\ref{eq:Lax(lambda)}) follows.

5. From $dt_{m\alpha} \wedge dt_{n\beta}$,
\beqn
    \dfrac{\rd ( \cU(\lambda),\cV(\lambda) )}
          {\rd ( t_{m\alpha}, t_{n\alpha} )}
    =   \dfrac{\rd \cB_{n\beta}(\lambda)}{\rd t_{m\alpha}}
      - \dfrac{\rd \cB_{m\alpha}(\lambda)}{\rd t_{n\beta}}.
\eeqn
The left hand side can be rewritten in the same way as above:
\beqnarray
    \dfrac{\rd\left( \cU(\lambda),\cV(\lambda) \right)}
          {\rd\left( t_{m\alpha}, t_{n\alpha} \right)}
    &=& \{ \cB_{m\alpha}(\lambda), \cU(\lambda) \}_\theta
           \{ \cB_{n\beta}(\lambda), \cV(\lambda) \}_\theta
                                                  \nonumber \\
    &&- \{ \cB_{m\alpha}(\lambda), \cV(\lambda) \}_\theta
           \{ \cB_{n\beta}(\lambda), \cU(\lambda) \}_\theta
                                                  \nonumber \\
    &=& \{ \cB_{m\alpha}(\lambda), \cB_{n\beta}(\lambda) \}_\theta.
\eeqnarray
Thus (\ref{eq:ZeroCurv(lambda)}) follows.

By carefully chasing these calculations, one will be able to show
that the converse is also true.

\section{Symmetries in Lax formalism}

\subsection{Main results}

There are several different ways to obtain additional symmetries
of this type of hierarchies.  The most orthodox method is due
to the twistor theoretical description of solutions
\cite{bib:Boyer-Plebanski}. The set of all twistor data $f$
forms a group of symplectic diffeomorphism, which acts on itself
by left or right multiplications.  Symmetries of the hierarchies
are nothing but the action of this group on the space of solutions
induced by the correspondence between twistor data and solutions.
Actually, this group action is highly nonlinear and difficut to
describe in any explicit form.  Its infinitesimal action, however,
turns out to have a very explicit and simple form.  This fact
was pointed out first in the case of the self-dual Einstein
equation \cite{bib:T-SDG}, and has been verified for some other
equations with a twistor theoretical structure.

Infinitesimal symmetries can also be obtained by directly solving
their defining equations, such as (\ref{eq:dKPSymmCond1}) in the
case of the dispersionless KP hierarchy.  In the present case,
these equations consist of the following:

\noindent (a) Four equations concerning Lax equations
(\ref{eq:Lax}),
\beqn
  \dfrac{\rd}{\rd t_{n\alpha}} \delta_\cA \cL
    = \{ \delta_\cA \cB_{n\alpha}, \cL \}
      + \{ \cB_{n\alpha}, \delta_\cA \cL \}
  \quad
    \Bigl( = \delta_\cA \dfrac{\rd \cL}{\rd t_{n\alpha}} \Bigr),
  \quad \mbox{etc.,}
                                     \label{eq:SymmCond1}
\eeqn
(replacing $\cL \to \cM,\cU,\cV$),
where $\delta_\cA \cB_{n\alpha}$ denotes the action on
$\cB_{n\alpha}$ induced by the relation $\cB_{n\alpha} =
(\cL^n e^{i\alpha\cU})_{\ge 0}$.  (These equations are
understood to include the equations with respect to $x$,
which are identical to the equations with respect to $t_{10}$.)

\noindent (b) Six equations concerning canonical relations
(\ref{eq:CanRel}),
\beqn
  \{ \delta_\cA \cL, \cM\} + \{ \cL, \delta_\cA \cM \} = 0,
  \quad \mbox{etc.,}
                                    \label{eq:SymmCond2}
\eeqn
(replacing $\cL,\cM \to \cL,\cM,\cU,\cV$).

If one can solve these equations for $\delta\cL$, etc. (and we
shall indeed give a large set of such solutions below), the
solutions give infinitesimal symmetries of the toroidal model.
A very successful example of this kind of direct approach
can be seen in the work of Park \cite{bib:Park-IJMP} on
symmetries of the self-dual Einstein equations and related
equations.

We now present our results on infinitesimal symmetries along
the line of second approach.  Actually, these results can
also be derived from the twistor theoretical appoach,
though we omit details.  We shall further calculate commutation
relations of those infinitesimal symmetries. Although the
results are formulated in the framework of the toroidal model,
basically the same construction can be applied to the planar
model (and presumably to any other models).

Let $\cR_0$ be the differential ring with generators
$(g_n,h_n,u_n,v_n)$ and derivations $(\rd/\rd x, \rd/\rd \theta_j,
\rd/\rd t_{n\alpha})$, and $R$ the extension explicitly including
the variables $(x,\theta_j,t_{n\alpha})$.
These differential rings represent the extended Lax system of
the toroidal model.

\begin{theorem} \label{th:Symm}
Given an arbitrary Laurent-Fourier series of the form
\beqn
    \cA(\lambda,\mu,u,v)
    = \sum_{n=-\infty}^\infty
        \sum_{m=0}^\infty
          \sum_{\alpha,\beta=-\infty}^\infty
            \lambda^n \mu^m e^{i(\alpha u + \beta v)},
\eeqn
let $\delta_\cA$ denote a new derivation $\delta_\cA: \cR \to \cR$
defined by
\beqnarray
    \delta_\cA \cL &=& \{ \cA(\cL,\cM,\cU,\cV)_{\le -1}, \cL \},
                                                  \nonumber \\
    \delta_\cA \cM &=& \{ \cA(\cL,\cM,\cU,\cV)_{\le -1}, \cM \},
                                                  \nonumber \\
    \delta_\cA \cU &=& \{ \cA(\cL,\cM,\cU,\cV)_{\le -1}, \cU \},
                                                  \nonumber \\
    \delta_\cA \cV &=& \{ \cA(\cL,\cM,\cU,\cV)_{\le -1}, \cV \},
                                                  \nonumber \\
    \delta_\cA t_{n\alpha}
       &=& \delta_\cA x = \delta_\cA \theta_j = 1.
                                                \label{eq:Symm}
\eeqnarray
The four quantities $(\delta_\cA \cL, \delta_\cA \cM,
\delta_\cA \cU, \delta_\cA \cV)$ satisfy (\ref{eq:SymmCond1})
and (\ref{eq:SymmCond2}), so that the derivation $\delta_\cA$
gives a symmetry of the toroidal model.
\end{theorem}

\begin{theorem} \label{th:CommRel}
These symmetries obey the commutation relations
\beqn
    [ \delta_\cA, \delta_\cB ]
       = \delta_{ \{\cA,\cB\}_{\lambda\mu uv} },
                                             \label{eq:CommRel}
\eeqn
where $\cA = \cA(\lambda,\mu,u,v)$ and $\cB = \cB(\lambda,\mu,u,v)$
are two arbitrary Laurent-Fourier series as above, and the Poisson
bracket of $\cA$ and $\cB$ is defined by
\beqn
    \{ \cA, \cB \}_{\lambda\mu uv}
    =   \frac{\rd \cA}{\rd \lambda} \frac{\rd \cB}{\rd \mu}
      - \frac{\rd \cA}{\rd \mu} \frac{\rd \cB}{\rd \lambda}
      + \frac{\rd \cA}{\rd u} \frac{\rd \cB}{\rd v}
      - \frac{\rd \cA}{\rd v} \frac{\rd \cB}{\rd u}.
\eeqn
\end{theorem}

Proof of these results are presented in the next two subsections.

\subsection{Proof of the first theorem}

We here prove Theorem \ref{th:Symm}. To simplify notations,
let $\cA$ denote $\cA(\cL,\cM,\cU,\cV)$ in the following.
We have to show that (\ref{eq:SymmCond1}) and
(\ref{eq:SymmCond2}) are indeed satisfied.

Let us examine (\ref{eq:SymmCond1}). By the defining equation
of $\delta_\cA \cL$ and the Lax equations, the left hand side of
(\ref{eq:SymmCond1}) can be calculated as:
\beqnarray
  \mbox{LHS of (\ref{eq:SymmCond1})}
  &=& \dfrac{\rd}{\rd t_{n\alpha}} \{ \cA_{\le -1}, \cL \}
                                                      \nonumber \\
  &=& \left\{ \dfrac{\rd \cA_{\le -1}}{\rd t_{n\alpha}}, \cL \right\}
    + \left\{ \cA_{\le -1}, \dfrac{\rd \cL}{\rd t_{n\alpha}} \right\}
                                                      \nonumber \\
  &=& \left\{ \dfrac{\rd \cA_{\le -1}}{\rd t_{n\alpha}}, \cL \right\}
    + \left\{ \cA_{\le -1}, \{ \cB_{n\alpha}, \cL\} \right\}.
\eeqnarray
Similarly,
\beqn
    \mbox{RHS of (\ref{eq:SymmCond1})}
    = \{ \delta_\cA \cB_{n\alpha}, \cL \}
      + \left\{ \cB_{n\alpha}, \{ \cA_{\le -1},
          \cB_{n\alpha} \} \right\}.
\eeqn
Subtracting the former from the latter and using the Jacobi identity
for the Poisson bracket, one can show that
\beqnarray
   \mbox{LHS} - \mbox{RHS} \ \mbox{of (\ref{eq:SymmCond1})}
   &=& \left\{
         \dfrac{\rd \cA_{\le -1}}{\rd t_{n\alpha}}
           - \delta_\cA \cB_{n\alpha}
             + \{ \cA_{\le -1}, \cB_{n\alpha} \},
               \cL \right\}.
                                           \label{eq:SymmCond1bis}
\eeqnarray

We now argue that
\beqn
         \dfrac{\rd \cA_{\le -1}}{\rd t_{n\alpha}}
           - \delta_\cA \cB_{n\alpha}
             + \{ \cA_{\le -1}, \cB_{n\alpha} \}
         = 0.
\eeqn
To show this relation, let us note the following obvious
consequence of the Lax equations and the Leibniz rule
satisfied by $\{\cB_{n\alpha},\cdot\}$:
\beqn
    \dfrac{\rd \cA}{\rd t_{n\alpha}} = \{ \cB_{n\alpha}, \cA \}.
\eeqn
The $(\quad)_{\le -1}$ part of this identity gives
\beqn
    \dfrac{\rd \cA_{\le -1}}{\rd t_{n\alpha}}
      = ( \{ \cB_{n\alpha}, \cA_{\le -1} \} )_{\le -1}.
\eeqn
Similarly, by the defining equation of $\delta_\cA$ and the
Leibniz rule satisfied by $\{ \cA_{\le -1}, \cdot \}$,
\beqn
    \delta_\cA \left( \cL^n e^{i\alpha\cU} \right)
      = \{ \cA_{\le -1}, \cL^n e^{i\alpha\cU} \}.
\eeqn
The $(\quad)_{\ge 0}$ part of both hand sides gives
\beqn
    \delta_\cA \cB_{n\alpha}
      = ( \{ \cA_{\le -1}, \cB_{n\alpha} \} )_{\ge 0}.
\eeqn
Combining these two equations, one finds that
\beqnarray
    \dfrac{\rd \cA_{\le -1}}{\rd t_{n\alpha}}
      - \delta_\cA \cB_{n\alpha}
    &=& ( \{ \cB_{n\alpha}, \cA_{\le -1} \} )_{\le -1}
      + ( \{ \cA_{\le -1}, \cB_{n\alpha} \} )_{\ge 0}
                                                       \nonumber \\
    &=& \{ \cB_{n\alpha}, \cA_{\le -1} \},
\eeqnarray
which is exactly the relation that we wish to derive.

Thus the right hand side of (\ref{eq:SymmCond1bis}) turns out to
vanish, and (\ref{eq:SymmCond1}) is proven for $\delta_\cA \cL$.
Other equations of (\ref{eq:SymmCond1}), too, can be checked by the
same way.

We next consider (\ref{eq:SymmCond2}). They are rather immediate
from the Jacobi identities. For instance,
\beqnarray
    \{ \delta_\cA \cL, \cM \} + \{ \cL, \delta_\cA \cM \}
    &=& \left\{ \{ \cA_{\le -1}, \cL \}, \cM \right\}
      + \left\{ \cL, \{ \cA_{\le -1}, \cM \} \right\}
                                                      \nonumber \\
    &=& \left\{ \cA_{\le -1}, \{ \cL, \cM \} \right\}
                                                      \nonumber \\
    &=& \{ \cA_{\le -1}, 1 \}
                                                      \nonumber \\
    &=& 0.
\eeqnarray
The other equations of (\ref{eq:SymmCond2}) can be derived in
the same way.

\subsection{Proof of the second theorem}

We now prove Theorem \ref{th:CommRel}.  Also here, $\cA$ and
$\cB$ denote $\cA(\cL,\cM,\cU,\cV)$ and $\cB(\cL,\cM,\cU,\cV)$.
As we shall show below, the proof is almost parallel to the proof
Theorem \ref{th:Symm}.

First examine the difference of both hand side of (\ref{eq:CommRel}).
One can calculate the repeated action $\delta_\cA \delta_\cB \cL$
of two symmetries just as in the calculation of the right hand
side of (\ref{eq:SymmCond1}):
\beqnarray
    \delta_\cA \delta_\cB \cL
    &=& \delta_\cA \{ \cB_{\le -1}, \cL \}     \nonumber \\
    &=& \{ \delta_\cA \cB_{\le -1}, \cL \}
      + \{ \cB_{\le -1}, \delta_\cA \cL \}     \nonumber \\
    &=& \{ \delta_{\le -1}, \cL\}
      + \left\{ \cB_{\le -1}, \{ \cA, \cL \} \right\}.
\eeqnarray
Subtracting from this the same equation with $\cA$ and $\cB$
interchanged, and using the Jacobi identity, one can derive
\beqn
    [\delta_\cA, \delta_\cB] \cL
      = \left\{
          \delta_\cA \cB_{\le -1}
            - \delta_\cB \cA_{\le -1}
              + \{ \cB_{\le -1}, \cA_{\le -1} \},
                \cL \right\}.
\eeqn
Meanwhile, by canonical relations (\ref{eq:CanRel}), one can
easily show that
\beqn
    \{ \cA, \cB \}
    = \left. \{ \cA(\lambda,\mu,u,v),
         \cB(\lambda,\mu,u,v) \}_{\lambda\mu uv}
            \right|_{\lambda=\cL,\mu=\cM,u=\cU,v=\cV}.
\eeqn
Combining this with the above identity, one obtains the relation
\beqnarray
   && [\delta_\cA, \delta_\cB] \cL
        - \delta_{ \{\cA,\cB\}_{\lambda\mu uv} } \cL
                                                     \nonumber \\
   &=& \left\{
          \delta_\cA \cB_{\le -1}
            - \delta_\cB \cA_{\le -1}
              + \{ \cB_{\le -1}, \cA_{\le -1} \}
                - (\{ \cA, \cB \}_{\lambda\mu uv})_{\le -1},
                  \cL \right\}.
                                     \label{eq:SymmCond2bis}
\eeqnarray

We now argue that
\beqn
          \delta_\cA \cB_{\le -1}
            - \delta_\cB \cA_{\le -1}
              + \{ \cB_{\le -1}, \cA_{\le -1} \}
                - (\{ \cA, \cB \}_{\lambda\mu uv})_{\le -1}
                  = 0.
\eeqn
To show this, we now start from the identities
\beqn
    \delta_\cA \cB = \{ \cA_{\le -1}, \cB \}, \quad
    \delta_\cB \cA = \{ \cB_{\le -1}, \cA \},
\eeqn
which follow from the definition of $\delta_\cA$ and $\delta_\cB$
and the Leibniz rule satisfied by $\delta_\cA$, $\delta_\cB$,
$\{\cA_{\le -1},\cdot\}$ and $\{\cB_{\le -1},\cdot\}$. Now one
can use the same method of calculations as employed in the proof
of Theorem \ref{th:Symm} to show that

\beqnarray
    & & \delta_\cA \cB_{\le -1}
          - \delta_\cB \cA_{\le -1}
            - \{ \cA_{\le -1}, \cB_{\le -1} \}
              - ( \{\cA, \cB\} )_{\le -1}  \nonumber \\
    &=& (
          \{ \cA_{\le -1}, \cB \}
            - \{ \cB_{\le -1}, \cA \}
              - \{ \cA_{\le -1}, \cB_{\le -1} \}
                - \{ \cA, \cB \}
                  )_{\le -1},                   \nonumber
\eeqnarray
and further substituting $\cA = \cA_{\le -1} + \cA_{\ge 0}$ etc.,
\beqnarray
    &=& (
          \{ \cA_{\le -1}, \cB_{\le -1} \}
            + \{ \cA_{\le -1}, \cB_{\ge 0} \}
              - \{ \cB_{\le -1}, \cA_{\le -1} \}
                - \{ \cB_{\le -1}, \cA_{\ge 0} \}   \nonumber \\
    &&  - \{ \cA_{\le -1}, \cB_{\le -1} \}          \nonumber \\
    &&  - \{ \cA_{\le -1}, \cB_{\le -1} \}
          - \{ \cA_{\le -1}, \cB_{\ge 0} \}
            - \{ \cA_{\ge 0}, \cB_{\le -1} \}
              - \{ \cA_{\ge 0}, \cB_{\ge 0} \}
                )_{\le -1}                          \nonumber \\
    &=& 0.
\eeqnarray
This is the relation that we wish to derive, which implies that
(\ref{eq:SymmCond2}) is satisfied by $\cL$ and $\cM$.

The other conditions of (\ref{eq:SymmCond2}), too, can be similarly
checked.

\subsection{Another expression of symmetries}

The following result, which we shall use later, shows what the
action of infinitesimal symmetries look like in terms of the
the Laurent series $\cM(\lambda)$, $\cU(\lambda)$ and
$\cV(\lambda)$.

\begin{theorem} \label{th:Symm(lambda)}
The symmetries $\delta_\cA$ act on $\cM(\lambda)$, $\cU(\lambda)$
and $\cV(\lambda)$ as:
\beqnarray
    \delta_\cA \cM(\lambda)
    &=& \dfrac{\rd \cA_{\le -1}(\lambda)}{\rd \lambda}
        + \{ \cA_{\le -1}(\lambda), \cM(\lambda) \}_\theta,
                                                \nonumber \\
    \delta_\cA \cU(\lambda)
    &=& \{ \cA_{\le -1}(\lambda), \cU(\lambda) \}_\theta,
                                                \nonumber \\
    \delta_\cA \cV(\lambda)
    &=& \{ \cA_{\le -1}(\lambda), \cV(\lambda) \}_\theta,
                                       \label{eq:Symm(lambda)}
\eeqnarray
where $\cA_{\le -1}(\lambda)$ is the Laurent series obtained
by expanding $\cA_{\le -1}$ into Laurent series of $\cL$,
\beqn
    \cA_{\le -1} = \sum_{n=-\infty}^\infty a_n(t,x,\theta) \cL^n,
\eeqn
and replacing $\cL \to \lambda$,
\beqn
    \cA_{\le -1}(\lambda)
       = \sum_{n=-\infty}^\infty a_n(t,x,\theta) \lambda^n.
\eeqn
\end{theorem}

\begin{proof}
Let us first derive the formula for $\cM(\lambda)$. Starting with
the obvious identity
\beqn
    \delta_\cA \cM
      = \delta_\cA \cM(\lambda)|_{\lambda=\cL}
        + \left.
            \dfrac{\rd \cM(\lambda)}{\rd \lambda}
              \right|_{\lambda=\cL}
                \delta_\cA \cL,
                                   \label{eq:Symm(lambda)Proof1}
\eeqn
we now calculate both hand sides as follows. First, by the
Leibniz rule satisfied by $\{\cdot,\cL\}$,
\beqnarray
    \delta_\cA \cL
    &=&  \{ \cA_{\le -1}, \cL \}
                                              \nonumber \\
    &=& \{ \cA_{\le -1}(\lambda), \cL \}|_{\lambda=\cL}
      + \left.
          \dfrac{\rd \cA_{\le -1}(\lambda)}{\rd \lambda}
            \right|_{\lambda=\cL}
              \{ \cL, \cL \},
\eeqnarray
and since $\{ \cL, \cL \} = 0$, the right hand side of
(\ref{eq:Symm(lambda)Proof1}) can be rewritten
\beqn
    \mbox{RHS of (\ref{eq:Symm(lambda)Proof1})}
      = \delta_\cA \cM(\lambda)|_{\lambda=\cL}
        + \left.
            \dfrac{\rd \cM(\lambda)}{\rd \lambda}
              \{ \cA_{\le -1}(\lambda), \cL \}
                \right|_{\lambda=\cL}.
\eeqn
Meanwhile, by the Leibniz rule satisfied by $\{\cdot,\cM\}$, similarly,
\beqnarray
    \delta_\cA \cM
      &=& \{\cA_{\le -1}, \cM \}
                                                \nonumber \\
      &=& \{ \cA_{\le -1}(\lambda), \cM \}|_{\lambda=\cL}
          + \left.
              \dfrac{\rd \cA_{\le -1}(\lambda)}{\rd \lambda}
                \right|_{\lambda=\cL}
                  \{ \cL, \cM \}.
\eeqnarray
Since $\{ \cL, \cM \} = 1$, and the Leibniz rule satisfied by
$\{\cA_{\le -1}(\lambda), \cdot\}$ implies the identity
\beqn
    \{ \cA_{\le -1}(\lambda), \cM \}
    = \left. \{ \cA_{\le -1}(\lambda), \cM(\lambda) \}
         \right|_{\lambda=\cL}
    + \left. \dfrac{\rd \cM(\lambda)}{\rd \lambda}
         \right|_{\lambda=\cL}
           \{ \cA_{\le -1}(\lambda), \cL \},
\eeqn
we now obtain the following expression for the left hand side of
(\ref{eq:Symm(lambda)Proof1}):
\beqnarray
    \mbox{LHS of (\ref{eq:Symm(lambda)Proof1})}
    &=& \{\cA_{\le -1}(\lambda),\cM(\lambda)\}|_{\lambda=\cL}
      + \left.
          \dfrac{\rd \cM(\lambda)}{\rd \lambda}
            \{ \cA_{\le -1}(\lambda), \cL \}
              \right|_{\lambda=\cL}
                                                    \nonumber \\
    && +\left.
          \dfrac{\rd \cA_{\le -1}(\lambda)}{\rd \lambda}
            \right|_{\lambda=\cL}.
\eeqnarray
Plugging these relations into (\ref{eq:Symm(lambda)Proof1}) gives
\beqn
    \delta_\cA \cM(\lambda)|_{\lambda=\cL}
    = \left.
        \dfrac{\rd \cA_{\le -1}(\lambda)}{\rd \lambda}
          \right|_{\lambda=\cL}
    + \{ \cA_{\le -1}(\lambda), \cM(\lambda) \}|_{\lambda=\cL}.
\eeqn
We can now remove ``$\lambda=\cL$'' to obtain
\beqn
    \delta_\cA \cM(\lambda)
    = \dfrac{\rd \cA_{\le -1}(\lambda)}{\rd \lambda}
      + \{ \cA_{\le -1}(\lambda), \cM(\lambda) \}.
\eeqn
The last Poisson bracket can be replaced by $\{\quad, \quad\}_\theta$
because both $\cA_{\le -1}(\lambda)$ and $\cM(\lambda)$ do not depend
on $k$.  The first formula of (\ref{eq:Symm(lambda)}) is thus proven.

The formulas for $\cU(\lambda)$ and $\cV(\lambda)$, too,  can be
derived in exactly the same way.  Now start from the identity
\beqn
    \delta_\cA \cU
    = \delta_\cA \cU(\lambda)|_{\lambda=\cL}
    + \left.
        \dfrac{\rd \cU(\lambda)}{\rd \lambda}
          \right|_{\lambda=\cL}
            \delta_\cA \cL,
                                             \label{eq:SymmProof2}
\eeqn
and calculate both hand sides. The right hand side can be calculated as:
\beqnarray
    \mbox{RHS of (\ref{eq:SymmProof2})}
    &=& \delta_\cA \cU(\lambda)|_{\lambda=\cL}
      + \left.
          \dfrac{\rd \cU(\lambda)}{\rd\lambda}
            \{ \cA_{\le -1}, \cL \}
              \right|_{\lambda=\cL}
                                                \nonumber \\
    &=& \delta_\cA \cU(\lambda)|_{\lambda=\cL}
      + \left.
          \dfrac{\rd \cU(\lambda)}{\rd \lambda}
            \{ \cA_{\le -1}(\lambda), \cL \}
              \right|_{\lambda=\cL}.
\eeqnarray
Similarly,
\beqnarray
    \mbox{LHS of (\ref{eq:SymmProof2})}
    &=& \{ \cA_{\le -1}, \cU \}
                                                \nonumber \\
    &=& \{ \cA_{\le -1}(\lambda), \cU \}|_{\lambda=\cL}
      + \{ \cA_{\le -1}(\lambda), \cL \}|_{\lambda=\cL}
          \{ \cL, \cU \},
                                                \nonumber \\
    &=& \{ \cA_{\le -1}(\lambda), \cU(\lambda) \}|_{\lambda=\cL}
      + \{ \cA_{\le -1}(\lambda), \cL \}
           \left.
              \dfrac{\rd \cU(\lambda)}{\rd \lambda}
                 \right|_{\lambda=\cL}.
\eeqnarray
Combining these relations, we now obtain
\beqn
    \delta_\cA \cU(\lambda)|_{\lambda=\cL}
    = \{ \cA_{\le -1}(\lambda), \cU(\lambda) \}|_{\lambda=\cL}.
\eeqn
Again ``$\lambda=\cL$'' can be removed, and the Poisson bracket in
the resulting equation may be replaced by $\{\quad,\quad\}_\theta$,
so that we can derive the formula for $\cU(\lambda)$.
The formula for $\cV(\lambda)$ can be treated in the same way.
\end{proof}

\section{$F$ function and symmetries}

\subsection{Main results}

We here introduce an analogue of the $F$ function for the toroidal
model of higher dimensional dispersionless hierarchy, and extend
the infinite dimensional symmetries of the last section to this
$F$ function. The extended symmetries turn out to obey anomalous
commutation relations.

The $F$ function of the dispersionless KP hierarchy is defined
by (\ref{eq:dKPDefEqF1}) (or by and (\ref{eq:dKPDefEqF2})).
Note that these equations can be rewritten
\beqnarray
    \dfrac{\rd F}{\rd t_n}
      &=& \res_\lambda \lambda^n \cM(\lambda)
          \quad (n = 1,2,\ldots),
                                               \nonumber \\
    \dfrac{\rd F}{\rd x}
      &=& \res_\lambda \lambda \cM(\lambda).
\eeqnarray
The following theorem shows a similar result for the toroidal
model.

\begin{theorem} \label{th:DefEqF}
Suppose that for any $\lambda$ in a neighborhood
of $\infty$, the symplectic map $\theta \mapsto
(\cU(\theta),\cV(\theta))$ is a diffeomorphism of the
torus onto itself. The following system of equations is then
integrable in the sense of Frobenius, and hence defines a
function $F = F(t,x)$ of $(t,x)$ up to an integration
constant:
\beqnarray
    \dfrac{\rd F}{\rd t_{n\alpha}}
      &=& \int \frac{d\theta_1 d\theta_2}{(2\pi)^2} \res_\lambda
            \lambda^n \cM(\lambda) e^{i\alpha\cU(\lambda)},
                                                   \nonumber \\
    \dfrac{\rd F}{\rd x}
      &=& \int \frac{d\theta_1 d\theta_2}{(2\pi)^2} \res_\lambda
            \lambda \cM(\lambda),
                                                   \nonumber \\
    \dfrac{\rd F}{\rd \theta_j}
      &=& 0,
                                              \label{eq:DefEqF}
\eeqnarray
where the integrals with respect to $\theta$ are over the torus
$T^2$ ($0 \le \theta_j \le 2\pi$).
\end{theorem}

\begin{remark}
1. Since the right hand side of the equations of $\rd F/\rd t_{10}$
and $\rd F/\rd x$ are the same, the $F$ function, like other
quantities, depends on $t_{10}$ and $x$ only through the linear
combination $t_{10} + x$. \newline
2. The $F$ function of the dispersionless dKP hierarchy can
reproduce all other quantities ($g_n$, $h_n$, etc.) by
differentiation. Because of this, the whole hierarchy, in
principle, can be converted into a system of differential
equations for $F$. This is not the case of the toroidal model
(and presumably of other higher dimensional integrable
hierarchies).
\end{remark}

To formulate results on symmetries, let us recall the
language of differential rings.  Following the case of the
dispersionless KP hierarchy, we enlarge the rings $\cR_0$ and
$\cR$ into $\cR_0[F]$ and $\cR[F]$. Eqs. (\ref{eq:DefEqF})
are then considered to be defining derivations
$\rd/\rd t_{n\alpha}$, $\rd/\rd x$ and $\rd/\rd \theta_j$
on these rings rather than defining $F$ itself, and the
theorem asserts that this extension of derivations is
well defined.

\begin{theorem} \label{th:extSymm}
The symmetries $\delta_\cA$ on $\cR$ can be extended to symmetries
on $\cR[F]$ by
\beqn
    \delta_\cA F
      = - \int \frac{d\theta_1 d\theta_2}{(2\pi)^2}
            \res_\lambda \int_0^{\cM(\lambda)} d\mu
                \cA( \lambda,\mu,\cU(\lambda),\cV(\lambda) ).
                                              \label{eq:SymmF}
\eeqn
Namely, this defines a new additional derivation on $\cR[F]$
that commutes with $\rd/\rd t_{n\alpha}$, $\rd/\rd x$ and
$\rd/\rd \theta_j$.
\end{theorem}

\begin{theorem} \label{th:extCommRel}
Under the same assumption as Theorem \ref{th:DefEqF}, these symmetries
on $\cR[F]$ obey the anomalous commutation relations
\beqn
    [\delta_\cA,\delta_\cB]
        = \delta_{ \{\cA,\cB\}_{\lambda\mu uv} }
          + c(\cA,\cB) \rd_F,
                                            \label{eq:extCommRel}
\eeqn
where $\{\quad,\quad\}_{\lambda\mu uv}$ is the same four dimensional
Poisson bracket as in Theorem \ref{th:CommRel}, $c(\cA,\cB)$ is
a cocycle of the four dimensional Poisson algebra given by
\beqn
    c(\cA,\cB)
      = \int \frac{dudv}{(2\pi)^2} \res_\lambda
            \cA(\lambda,0,u,v)
               \frac{\rd}{\rd \lambda}
                  \cB(\lambda,0,u,v),
\eeqn
and $\rd_F$ is another derivation on $\cR[F]$ uniquely determined by
\beqn
    \rd_F (F) = 1, \quad
    \rd_F (\mbox{any other generator of $\cR$}) = 0.
\eeqn
\end{theorem}

Thus the commutation relations of extended symmetries obet the
structure of a one-dimensional central extension of the four
dimensional Poisson algebra with the Poisson bracket
$\{ \cA,\cB \}_{\lambda\mu uv}$. This result, too, is a
very natural extension of the corresponding result in
the dispersionless KP hierarchy.

The proofs of these theorems are more or less similar to each other.
We first present the proof of Theorem \ref{th:DefEqF} in some detail,
and then organize the others in much the same, but slightly compressed
form.

\subsection{Proof of the first theorem}

We now prove Theorem \ref{th:DefEqF}. Let $h_{n\alpha}$
$h_{10}$ denote the right hand side of the first equation of
(\ref{eq:DefEqF}):
\beqn
    h_{n\alpha}
    = \int \frac{d\theta_1 d\theta_2}{(2\pi)^2} \res_\lambda
        \lambda^n \cM(\lambda) e^{i\alpha\cU(\lambda)}.
\eeqn
The Frobenius integrability condition for these quantities are
given by
\beqn
    \dfrac{\rd h_{n\beta}}{\rd t_{m\alpha}}
      = \dfrac{\rd h_{m\alpha}}{\rd t_{n\beta}}.
                                       \label{eq:DefEqFProof1}
\eeqn
We have only to chech these conditions.  The others concerning
the $x$-derivatives are identical to those of $t_{10}$-derivatives,
and included in the above conditions as a special case.

1. Let us calculate $\rd h_{n\beta}/\rd t_{m\alpha}$. First,
differentiating the above defining equation of $h_{n\beta}$,
we have
\beqnarray
    \dfrac{\rd h_{n\beta}}{\rd t_{m\alpha}}
    &=& \int \frac{d\theta_1 d\theta_2}{(2\pi)^2} \res_\lambda
         \lambda^n
          \left(
             \dfrac{\rd e^{i\beta\cU(\lambda)}}{\rd t_{m\alpha}}
               \cM(\lambda)
           + e^{i\beta\cU(\lambda)}
               \dfrac{\rd \cM(\lambda)}{\rd t_{m\alpha}}
          \right),
                                                  \nonumber
\eeqnarray
and plugging Lax equations (\ref{eq:Lax(lambda)}) for $\cM(\lambda)$
and $\cU(\lambda)$,
\beqnarray
    &=& \int \frac{d\theta_1 d\theta_2}{(2\pi)^2} \res_\lambda
          \lambda^n
             \{ \cB_{m\alpha}(\lambda),
                  e^{i\beta\cU(\lambda)} \}_\theta
                    \cM(\lambda)
                                                        \nonumber \\
    &&+ \int \frac{d\theta_1 d\theta_2}{(2\pi)^2} \res_\lambda
          \lambda^n
             e^{i\beta\cU(\lambda)}
               \left(
                  \dfrac{\rd \cB_{m\alpha}(\lambda)}{\rd \lambda}
                    + \{ \cB_{m\alpha}(\lambda),\cM(\lambda) \}_\theta
                      \right)
                                                        \nonumber \\
    &=& \int \frac{d\theta_1 d\theta_2}{(2\pi)^2} \res_\lambda
          \lambda^n
            \{ \cB_{m\alpha}(\lambda),
               e^{i\beta\cU(\lambda)}\cM(\lambda) \}_\theta
                                                        \nonumber \\
    &&+ \int \frac{d\theta_1 d\theta_2}{(2\pi)^2} \res_\lambda
          \lambda^n
              e^{i\beta\cU(\lambda)}
                \dfrac{\rd \cB_{m\alpha}(\lambda)}{\rd \lambda}.
\eeqnarray
Let us examine the first and second terms in the last two lines:

(i) The first term turns out vanish because of the following lemma.

\begin{lemma} \label{lem:vanishing}
For any two functions $A = A(\theta)$ and $B = B(\theta)$ on the
torus,
\beqn
    \int \frac{d\theta_1 d\theta_2}{(2\pi)^2} \{ A, B \}_\theta = 0.
\eeqn
\end{lemma}

\begin{proof}
The Poisson bracket can be re written into a total derivative:
\beqn
    \{ A, B \}_\theta
    = \frac{1}{2}\dfrac{\rd}{\rd \theta_2}
        \left( \dfrac{\rd A}{\rd \theta_1}
               - A \dfrac{\rd B}{\rd \theta_1} \right)
    - \frac{1}{2}\dfrac{\rd}{\rd \theta_1}
        \left( \dfrac{\rd A}{\rd \theta_2}
               - A \dfrac{\rd B}{\rd \theta_2} \right).
\eeqn
\end{proof}

(ii) To handle the second term, recall the relation between
$\cB_{m\alpha}$ and $\cB_{m\alpha}(\lambda)$. Since $\cB_{m\alpha}$
depends on $k$ only through $\lambda = \cL(t,x,\theta,k)$,
\beqn
    \dfrac{\rd \cB_{m\alpha}}{\rd k}
    = \left. \dfrac{\rd \cB_{m\alpha}(\lambda)}{\rd \lambda}
      \right|_{\lambda = \cL}
      \dfrac{\rd \cL}{\rd k}.
\eeqn
By this relation and the following lemma, we can rewrite the
$\lambda$-residue into a $k$-residue:
\beqnarray
    \res_\lambda
       \lambda^n e^{i\beta\cU(\lambda)}
           \dfrac{\rd \cB_{m\alpha}(\lambda)}{\rd \lambda}
    &=& \res_k
          \cL^n e^{i\beta\cU}
            \left.
              \dfrac{\rd \cB_{m\alpha}(\lambda)}{\rd \lambda}
                \right|_{\lambda=\cL}
                  \dfrac{\rd \cL}{\rd k}
                                                \nonumber \\
    &=& \res_k
            \cL^n e^{i\beta\cU}
              \dfrac{\rd \cB_{m\alpha}}{\rd k}.
\eeqnarray

\begin{lemma} \label{lem:substitution}
Under a change of coordinates at $\infty$ of the form
$\lambda = \cL(k) = k + O(k^{-1})$, the residues with respect
to $\lambda$ and $k$ are connected as
\beqn
    \res_\lambda f(\lambda)
    = \res_k f(\cL) \frac{\rd \cL}{\rd k}.
\eeqn
\end{lemma}

\begin{proof}
Since the residue can be written as a contour integral (along
a small loop encircling $\infty$),
\beqn
    \res_\lambda f(\lambda)
    = \oint \frac{d\lambda}{2\pi i} f(\lambda),
\eeqn
the above formula is nothing but the formula of change of variables
of integrals, $\rd \cL/\rd k$ being the Jacobian.
\end{proof}

Plugging these into the previous expression of
$\rd h_{n\beta}/\rd t_{m\alpha}$, we arrive at the following
intermediate result:
\beqnarray
    \dfrac{\rd h_{n\beta}}{\rd t_{m\alpha}}
    &=& \int \frac{d\theta_1 d\theta_2}{(2\pi)^2} \res_k
          \cL^n e^{i\beta\cU} \frac{\rd}{\rd k} \cB_{m\alpha}
                                                \nonumber \\
    &=& \int \frac{d\theta_1 d\theta_2}{(2\pi)^2} \res_k
          \left( \cL^n e^{i\beta\cU} \right)_{\le -1}
            \frac{\rd}{\rd k}
              \left( \cL^m e^{i\alpha\cU} \right)_{\ge 0}.
\eeqnarray

2. We now subtract from the last equation the same equation
with $(m,\alpha)$ and $(n,\beta)$ interchanged. This gives
\beqnarray
    \dfrac{\rd h_{n\beta}}{\rd t_{m\alpha}}
    - \dfrac{\rd h_{m\alpha}}{\rd t_{n\beta}}
  &=& \int \frac{d\theta_1 d\theta_2}{(2\pi)^2} \res_k
      \left(
        \left( \cL^n e^{i\beta\cU} \right)_{\le -1}
          \frac{\rd}{\rd k}
            \left( \cL^m e^{i\alpha\cU} \right)_{\ge 0}
              \right.
                                                \nonumber \\
  && \qquad
         \left.
           - \left( \cL^m e^{i\alpha\cU} \right)_{\le -1}
               \frac{\rd}{\rd k}
                 \left( \cL^n e^{i\beta\cU} \right)_{\ge 0}
                   \right)
                                                \nonumber \\
  &=& \int \frac{d\theta_1 d\theta_2}{(2\pi)^2} \res_k
      \left(
        \left( \cL^n e^{i\beta\cU} \right)_{\le -1}
          \frac{\rd}{\rd k}
            \left( \cL^m e^{i\alpha\cU} \right)_{\ge 0}
              \right.
                                                \nonumber \\
  && \qquad
          \left.
            + \left( \cL^n e^{i\beta\cU} \right)_{\ge 0}
                \frac{\rd}{\rd k}
                  \left( \cL^n e^{i\beta\cU} \right)_{\le -1}
                    \right).
\eeqnarray
In the last line, the following lemma has been used.

\begin{lemma} \label{lem:antisymmetry}
\beqn
    \res_k f \frac{\rd g}{\rd k}
    = - \res_k g \frac{\rd f}{\rd k}.
\eeqn
\end{lemma}

\begin{proof}
Written in the language of contour integrals, this is nothing
buth the formula of integration by parts.
\end{proof}

\noindent Adding the obvious identities
\beqnarray
    \res_k
      \left( \cL^n e^{i\beta\cU} \right)_{\le -1}
        \frac{\rd}{\rd k}
          \left( \cL^m e^{i\alpha\cU} \right)_{\le -1}
    &=& 0,
                                            \nonumber \\
    \res_k
      \left( \cL^n e^{i\beta\cU} \right)_{\ge 0}
        \frac{\rd}{\rd k}
          \left( \cL^m e^{i\alpha\cU} \right)_{\ge 0}
    &=& 0
\eeqnarray
to the above result, we can rewrite it into a more compact form,
\beqnarray
    \dfrac{\rd h_{n\beta}}{\rd t_{m\alpha}}
    - \dfrac{\rd h_{m\alpha}}{\rd t_{n\beta}}
    &=& \int \frac{d\theta_1 d\theta_2}{(2\pi)^2} \res_k
          \left( \cL^n e^{i\beta\cU} \right)
            \frac{\rd}{\rd k}
              \left( \cL^m e^{i\alpha\cU} \right),
                                                \nonumber
\eeqnarray
and further using Lemma \ref{lem:substitution}, but now in the
inverse way, we can recast this into a $\lambda$-residue:
\beqn
   = \int \frac{d\theta_1 d\theta_2}{(2\pi)^2} \res_\lambda
         a( \lambda,\cU(\lambda),\cV(\lambda) )
           \frac{\rd}{\rd \lambda}
             b( \lambda,\cU(\lambda),\cV(\lambda) ).
                                           \label{eq:DefEqFProof2}
\eeqn

3. We now have to evaluate the last integral. Since a similar issue
also emerges in the proof of Theorem \ref{th:extCommRel}, let us
formulate it in a somewhat general form as follows:

\begin{lemma} \label{lem:evaluation}
Under the same assumption as in Theorem \ref{th:DefEqF}, and for
any Laurent-Fourier series $a(\lambda,u,v)$ and $b(\lambda,u,v)$
independent of $\mu$,
\beqnarray
  && \int \frac{d\theta_1 d\theta_2}{(2\pi)^2} \res_\lambda
         a( \lambda,\cU(\lambda),\cV(\lambda) )
           \frac{\rd}{\rd \lambda}
             b( \lambda,\cU(\lambda),\cV(\lambda) )
                                                \nonumber \\
  &=&  \int \frac{dudv}{(2\pi)^2} \res_\lambda
         a(\lambda,u,v) \frac{\rd}{\rd \lambda} b(\lambda,u,v)
                                                \nonumber \\
  && - \int \frac{d\theta_1 d\theta_2}{(2\pi)^2} \res_\lambda
         \cM(\lambda)
           \{ a(\lambda,u,v),b(\lambda,u,v) \}_\theta.
                                       \label{eq:evaluation}
\eeqnarray
\end{lemma}

\begin{proof}
By the Leibniz rule, the integrand on the left hand side of
(\ref{eq:evaluation}) can be expanded into several pieces:
\beqnarray
    &&  a( \lambda,\cU(\lambda),\cV(\lambda) )
          \frac{\rd}{\rd \lambda}
            b( \lambda,\cU(\lambda),\cV(\lambda) )
                                                   \nonumber \\
    &=& a( \lambda,\cU(\lambda),\cV(\lambda) )
           \left.\dfrac{\rd b(\lambda,u,v)}{\rd \lambda}
               \right|_{u=\cU(\lambda),v=\cV(\lambda)}
                                                   \nonumber \\
    && + a(  \lambda,\cU(\lambda),\cV(\lambda) )
           \left(
             \dfrac{\rd b(\lambda,u,v)}{\rd \cU(\lambda)}
               \dfrac{\rd \cU(\lambda)}{\rd \lambda}
                 + \dfrac{\rd b(\lambda,u,v)}{\rd \cV(\lambda)}
                   \dfrac{\rd \cV(\lambda)}{\rd \lambda}
                     \right).
                                    \label{eq:evaluationProof}
\eeqnarray
Contribution from the first part in the last two lines is given by
\beqnarray
  && \int \frac{d\theta_1 d\theta_2}{(2\pi)^2} \res_\lambda
        a(\lambda,u,v)
           \left.\frac{\rd b(\lambda,u,v)}{\rd \lambda}
             \right|_{u=\cU(\lambda),v=\cV(\lambda)}
                                                      \nonumber \\
  &=& \int \frac{dudv}{(2\pi)^2} \res_\lambda
        a(\lambda,u,v) \frac{\rd b(\lambda,u,v)}{\rd \lambda},
\eeqnarray
where, by the topological assumption, we have changed the
integration variables as $(\theta_1,\theta_2) \to (u,v)
= \left(\cU(\lambda),\cV(\lambda)\right)$. This gives the first
term on the right hand side of (\ref{eq:evaluation}). To evaluate
the contribution from the second part, we note the identity
\beqnarray
   && \{ \cM(\lambda),
          b( \lambda,\cU(\lambda),\cV(\lambda) ) \}_\theta
                                                  \nonumber \\
   \qquad &=&
     \dfrac{\rd b(\lambda,u,v)}{\rd \cU(\lambda)}
       \dfrac{\rd \cU(\lambda)}{\rd \lambda}
         + \dfrac{\rd b(\lambda,u,v)}{\rd \cV(\lambda)}
            \dfrac{\rd \cV(\lambda)}{\rd \lambda},
\eeqnarray
which one can easily show by (\ref{eq:CanRel(lambda)}).
Contribution from the second part of (\ref{eq:evaluationProof})
is thus given by the integral
\beqnarray
 &&  \int \frac{d\theta_1 d\theta_2}{(2\pi)^2} \res_\lambda
         a(\lambda,\cU(\lambda),\cV(\lambda) )
           \{ \cM(\lambda),
              b( \lambda,\cU(\lambda),\cV(\lambda) ) \}_\theta,
                                                  \nonumber
\eeqnarray
which can be rewritten
\beqnarray
   &=& \int \frac{d\theta_1 d\theta_2}{(2\pi)^2} \res_\lambda
           \{ \cM(\lambda) a( \lambda,\cU(\lambda),\cV(\lambda) ),
              b( \lambda,\cU(\lambda),\cV(\lambda) ) \}_\theta
                                                      \nonumber \\
   && - \int \frac{d\theta_1 d\theta_2}{(2\pi)^2} \res_\lambda
           \cM(\lambda)
             \{ a( \lambda,\cU(\lambda),\cV(\lambda) ),
                b( \lambda,\cU(\lambda),\cV(\lambda) ) \}_\theta.
\eeqnarray
The first term first term vanishes by Lemma \ref{lem:vanishing},
and the second term gives the second term on the right hand side
of (\ref{eq:evaluation}). Thus (\ref{eq:evaluation}) is proven.
\end{proof}

Returning to the case under consideration, $a$ and $b$ are given by
\beqn
    a(\lambda,u,v) = \lambda^m e^{i\alpha u}, \quad
    b(\lambda,u,v) = \lambda^n e^{i\beta u},
\eeqn
Applying the above lemma to this case, we can readily see that
the right hand side of (\ref{eq:DefEqFProof2}) vanishes, so that
(\ref{eq:DefEqFProof1}) are indeed satisfied.

\subsection{Proof of the second theorem}

To prove Theorem \ref{th:extSymm}, we have to show that
\beqn
    \delta_\cA \dfrac{\rd F}{\rd t_{m\alpha}}
    = \dfrac{\rd}{\rd t_{m\alpha}} \delta_\cA F.
                                      \label{eq:extSymmProof1}
\eeqn
We calculate both hand sides separately, and show that they
coincide. The method of calculations are almost parallel to
the last subsection, simply $\rd/\rd t_{n\beta}$ being
replaced by $\delta_\cA$.

1. Let us consider the left hand side of (\ref{eq:extSymmProof1}).
This is similar to the calculation of
$\rd h_{m\alpha} / \rd t_{n\beta}$ in the last subsection.
First, since $\delta_\cA$ (like $\rd/\rd t_{n\beta}$) is
also a derivation ,
\beqnarray
    \delta_\cA \dfrac{\rd F}{\rd t_{m\alpha}}
    &=& \delta_\cA
          \int \frac{d\theta_1 d\theta_2}{(2\pi)^2} \res_\lambda
            \lambda^m e^{i\alpha\cU(\lambda)} \cM(\lambda)
                                                        \nonumber \\
    &=& \int \frac{d\theta_1 d\theta_2}{(2\pi)^2} \res_\lambda
          \lambda^m
            \left(
              \delta_\cA e^{i\alpha\cU(\lambda)} \cdot \cM(\lambda)
              + e^{i\alpha\cU(\lambda)} \cdot \delta_\cA \cM(\lambda)
                  \right),
                                                        \nonumber
\eeqnarray
and evaluating the terms containing $\delta_\cA$ by using the defining
equation (\ref{eq:Symm(lambda)}) and the Leibniz rule of $\delta_\cA$,
we find that
\beqnarray
    &=& \int \frac{d\theta_1 d\theta_2}{(2\pi)^2} \res_\lambda
          \lambda^m
            \{ \cA_{\le -1}(\lambda), e^{i\alpha\cU(\lambda)} \}_\theta
                                                        \nonumber \\
    && + \int \frac{d\theta_1 d\theta_2}{(2\pi)^2} \res_\lambda
          \lambda^m
            e^{i\alpha\cU(\lambda)}
              \frac{\rd \cA_{\le -1}(\lambda)}{\rd \lambda}.
\eeqnarray
The first term in the last two lines vanish because of Lemma
\ref{lem:vanishing}. The second term can be rewritten into
a $k$-residue by means of Lemma \ref{lem:substitution}. Thus,
\beqnarray
    \delta_\cA \dfrac{\rd F}{\rd t_{m\alpha}}
    &=& \int \frac{d\theta_1 d\theta_2}{(2\pi)^2} \res_k
          \cL^m e^{i\alpha\cU} \frac{\rd}{\rd k} \cA_{\le -1}
                                                        \nonumber \\
    &=& \int \frac{d\theta_1 d\theta_2}{(2\pi)^2} \res_k
          \cB_{m\alpha} \frac{\rd}{\rd k} \cA_{\le -1}
                                                        \nonumber \\
    &=& - \int \frac{d\theta_1 d\theta_2}{(2\pi)^2} \res_k
          \cA_{\le -1} \frac{\rd}{\rd k} \cB_{m\alpha}.
                                          \label{eq:extSymmProof2}
\eeqnarray
(Lemma \ref{lem:antisymmetry}) has been used in deriving the
last line.)

2. Let us now consider the right hand side of
(\ref{eq:extSymmProof1}). Differentiating the defining equation
(\ref{eq:SymmF}) of $\delta_\cA F$, we first obtain
\beqnarray
    \dfrac{\rd}{\rd t_{m\alpha}} \delta_\cA F
    &=& - \dfrac{\rd}{\rd t_{m\alpha}}
              \int \frac{d\theta_1 d\theta_2}{(2\pi)^2}
                \res_\lambda \int_0^{\cM(\lambda)} d\mu
                  \cA( \lambda,\mu,\cU(\lambda),\cV(\lambda) )
                                                        \nonumber \\
    &=& - \int \frac{d\theta_1 d\theta_2}{(2\pi)^2} \res_\lambda
            \cA( \lambda,\cM(\lambda),\cU(\lambda),\cV(\lambda) )
              \dfrac{\rd \cM(\lambda)}{\rd t_{m\lambda}}
                                                        \nonumber \\
    &&  - \int \frac{d\theta_1 d\theta_2}{(2\pi)^2} \res_\lambda
            \int_0^{\cM(\lambda)} d\mu
              \frac{\rd}{\rd t_{m\alpha}}
                \cA( \lambda,\mu,\cU(\lambda),\cV(\lambda) ),
                                                        \nonumber
\eeqnarray
and using (\ref{eq:Lax(lambda)}) along with the chain rule of
differentiation,
\beqnarray
  &=& - \int \frac{d\theta_1 d\theta_2}{(2\pi)^2} \res_\lambda
         \cA( \lambda,\cM(\lambda),\cU(\lambda),\cV(\lambda) )
           \dfrac{\rd}{\rd \lambda} \cB_{m\alpha}(\lambda)
                                                      \nonumber \\
  &&  - \int \frac{d\theta_1 d\theta_2}{(2\pi)^2} \res_\lambda
            \cA( \lambda,\cM(\lambda),\cU(\lambda),\cV(\lambda) )
              \{ \cB_{m\alpha}(\lambda),\cM(\lambda) \}_\theta
                                                      \nonumber \\
  &&  - \int \frac{d\theta_1 d\theta_2}{(2\pi)^2} \res_\lambda
          \int_0^{\cM(\lambda)} d\mu
            \dfrac{\rd \cA}{\rd \cU(\lambda)}
              \{ \cB_{m\alpha}(\lambda), \cU(\lambda) \}_\theta
                                                      \nonumber \\
  &&  - \int \frac{d\theta_1 d\theta_2}{(2\pi)^2} \res_\lambda
          \int_0^{\cM(\lambda)} d\mu
            \dfrac{\rd \cA}{\rd \cV(\lambda)}
              \{ \cB_{m\alpha}(\lambda), \cV(\lambda) \}_\theta.
\eeqnarray
Then by the same reasoning as used in the last subsection,
the first term in the last four lines can be converted into
a $k$-residue of the form
\beqn
    - \int \frac{d\theta_1 d\theta_2}{(2\pi)^2} \res_k
        \cA \frac{\rd}{\rd k} \cB_{m\alpha}
    = - \int \frac{d\theta_1 d\theta_2}{(2\pi)^2} \res_k
          \cA_{\le -1} \frac{\rd}{\rd k} \cB_{m\alpha}.
\eeqn
The remaining part can be written as an integral of a single
Poisson bracket of the form (which vanishes by Lemma
\ref{lem:vanishing}):
\beqn
    - \int \frac{d\theta_1 d\theta_2}{(2\pi)^2} \res_\lambda
       \left\{ \cB_{m\alpha}(\lambda),
           \int_0^{\cM(\lambda)} d\mu
             \cA( \lambda,\mu,\cU(\lambda),\cV(\lambda) ) \right\}
    = 0.
\eeqn
Therefore,
\beqn
    \dfrac{\rd}{\rd t_{m\alpha}} \delta_\cA F
    = - \int \frac{d\theta_1 d\theta_2}{(2\pi)^2} \res_k
          \cA_{\le -1} \frac{\rd}{\rd k} \cB_{m\alpha}.
                                        \label{eq:extSymmProof3}
\eeqn

Thus (\ref{eq:extSymmProof2}) and (\ref{eq:extSymmProof3})
turn out to give an identical result, completing the proof
of (\ref{eq:extSymmProof1}).

\subsection{Proof of the third theorem}

We now prove Theorem \ref{th:extCommRel}. It is sufficient to
prove the special relation
\beqn
    [\delta_\cA,\delta_\cB] F
      = \delta_{\{ \cA,\cB \}_{\lambda\mu uv}} F + c(\cA,\cB).
                                         \label{eq:extCommRelProof1}
\eeqn
The following proof resembles the proof of Theorem \ref{th:DefEqF}
(and, indeed, includes it as a special case!).

1. We first calculate $\delta_\cA \delta_\cB F$. Details are
similar to the previous calculations. First, since $\delta_\cA$
(like $\rd/\rd t_{m\alpha}$) is also a derivation, its action on
the integral representation of $\delta_\cB F$ can be evaluated as:
\beqnarray
    \delta_\cA \delta_\cB F
    &=& - \delta_\cA
            \int \frac{d\theta_1 d\theta_2}{(2\pi)^2} \res_\lambda
              \int_0^{\cM(\lambda)} d\mu
                \cB( \lambda,\mu,
                          \cU(\lambda),\cV(\lambda) )
                                                    \nonumber \\
    &=& - \int \frac{d\theta_1 d\theta_2}{(2\pi)^2} \res_\lambda
             \cB( \lambda,\cM(\lambda),\cU(\lambda),\cV(\lambda) )
                \delta_\cA \cM(\lambda)
                                                    \nonumber \\
    &=& - \int \frac{d\theta_1 d\theta_2}{(2\pi)^2} \res_\lambda
             \int_0^{\cM(\lambda)} d\mu
               \delta_\cA
                 \cB( \lambda,\mu,\cU(\lambda),\cV(\lambda) ).
\eeqnarray
Second, using the defining equation (\ref{eq:SymmF}) and
the Leibniz rule of $\delta_\cA$ again, we can rewrite the above
formula into the following form:
\beqnarray
    \delta_\cA \delta_\cB F
    &=& - \int \frac{d\theta_1 d\theta_2}{(2\pi)^2} \res_\lambda
            \left\{ \cA_{\le -1}(\lambda),
              \int_0^{\cM(\lambda)} d\mu
                \cB( \lambda,\mu,\cU(\lambda),\cV(\lambda) )
                  \right\}_\theta
                                                    \nonumber \\
    &&  - \int \frac{d\theta_1 d\theta_2}{(2\pi)^2} \res_\lambda
            \cB( \lambda,\cM(\lambda),\cU(\lambda),\cV(\lambda) )
               \dfrac{\rd \cA_{\le -1}(\lambda)}{\rd \lambda}.
\eeqnarray
Now by the same reasoning as in the previous proofs, the first term
vanishes and the second term can be rewritten into a $k$-residue.
Thus we obtain the following expression of $\delta_\cA \delta_\cB F$:
\beqn
    \delta_\cA \delta_\cB F
    = - \int \frac{d\theta_1 d\theta_2}{(2\pi)^2} \res_k
          \cB(\cL,\cM,\cU,\cV)_{\ge 0}
            \frac{\rd}{\rd k}
              \cA(\cL,\cM,\cU,\cV)_{\le -1}.
\eeqn

2. Now Subtract from the last equation the same equation with
$\cA$ and $\cB$ interchanged. Also use Lemma \ref{lem:antisymmetry}
to reverse the order of the two terms sandwitching $\rd/\rd k$.
By basically the same calculations as in the preceding sections,
we eventually obtain
\beqnarray
   [\delta_\cA,\delta_\cB] F
    &=&  \int \frac{d\theta_1 d\theta_2}{(2\pi)^2} \res_k
         \cA(\cL,\cM,\cU,\cV)
           \frac{\rd}{\rd k}
             \cB(\cL,\cM,\cU,\cV),
                                                 \nonumber
\eeqnarray
and further by Lemma \ref{lem:substitution}, this can be converted
into a $\lambda$-residue, i.e.,
\beqn
    =  \int \frac{d\theta_1 d\theta_2}{(2\pi)^2} \res_\lambda
          \cA( \lambda,\cM(\lambda),\cU(\lambda),\cV(\lambda) )
            \frac{\rd}{\rd \lambda}
              \cB( \lambda,\cM(\lambda),\cU(\lambda),\cV(\lambda) ).
                                      \label{eq:extCommRelProof2}
\eeqn

3. We now have to evaluate the last line of (\ref{eq:extCommRelProof2}).
Since target equation (\ref{eq:extCommRelProof1}) is bilinear in
$\cA$ and $\cB$, it is sufficient to consider the case where
$\cA$ and $\cB$ take the following somewhat special form:
\beqn
    \cA = a(\lambda,u,v) \mu^j, \quad
    \cB = b(\lambda,u,v) \mu^k \quad (j,k \ge 0).
\eeqn

(i) {\it The case where $j + k = 0$ (i.e., $j = k = 0$)}:
In this case, the integrand of the last line of
(\ref{eq:extCommRelProof2}) takes exactly the same for as in
Lemma \ref{lem:evaluation}. The first term on the right hand
side of (\ref{eq:evaluation}) now gives the cocycle $c(\cA,\cB)$:
\beqn
     \int \frac{dudv}{(2\pi)^2} \res_\lambda
       a(\lambda,u,v) \frac{\rd}{\rd \lambda} b(\lambda,u,v)
     = c(a,b).
\eeqn
The last term of (\ref{eq:evaluation}), meanwhile, gives
$\delta_{\{a,b\}_{\lambda\mu uv}} F$:
\beqnarray
  && - \int \frac{d\theta_1 d\theta_2}{(2\pi)^2} \res_\lambda
         \cM(\lambda)
           \{ a( \lambda,\cU(\lambda),\cV(\lambda) ),
              b( \lambda,\cU(\lambda),\cV(\lambda) ) \}_\theta
                                                   \nonumber \\
  &=& - \int \frac{d\theta_1 d\theta_2}{(2\pi)^2} \res_\lambda
         \int_0^{\cM(\lambda)} d\mu
           \{ a( \lambda,\cU(\lambda),\cV(\lambda) ),
              b( \lambda,\cU(\lambda),\cV(\lambda) ) \}_\theta
                                                    \nonumber \\
  &=& \delta_{\{a,b\}_{\lambda\mu uv}} F.
\eeqnarray
(In the last line, we we have used the identity
\beqn
       \{ a( \lambda,\cU(\lambda),\cV(\lambda) ),
          b( \lambda,\cU(\lambda),\cV(\lambda) ) \}_\theta
       = \{a,b\}_{\lambda\mu uv}|_{u=\cU(\lambda),v=\cV(\lambda)},
\eeqn
which follows from the first equation of (\ref{eq:CanRel(lambda)}).)
Collecting these pieces, one can see that (\ref{eq:extCommRelProof1})
is indeed satisfied in this case.

(ii) {\it The case where $j + k > 0$}:
Now the integrand of (\ref{eq:extCommRelProof2}) becomes
\beqnarray
  &&  \cA( \lambda,\cM(\lambda),\cU(\lambda),\cV(\lambda) )
      \frac{\rd}{\rd \lambda}
      \cB( \lambda,\cM(\lambda),\cU(\lambda),\cV(\lambda) )
                                                         \nonumber \\
  &=& k \cM(\lambda)^{j+k-1} \frac{\rd \cM(\lambda)}{\rd \lambda}
      a( \lambda,\cU(\lambda),\cV(\lambda) )
      b( \lambda,\cU(\lambda),\cV(\lambda) )
                                                         \nonumber \\
  &&+ \cM(\lambda)^{j+k}
      a( \lambda,\cU(\lambda),\cV(\lambda) )
      \frac{\rd}{\rd \lambda}
      b( \lambda,\cU(\lambda),\cV(\lambda) )
                                                         \nonumber \\
  &=& \frac{\rd}{\rd \lambda}
        \left( \frac{k}{j+k} \cM(\lambda)^{j+k}
          a( \cU(\lambda),\cV(\lambda) )
            b( \cU(\lambda),\cV(\lambda) ) \right)
                                                         \nonumber \\
  &&+ \frac{j}{j+k} \cM(\lambda)^{j+k}
      a( \lambda,\cU(\lambda),\cV(\lambda) )
      \frac{\rd}{\rd \lambda}
      b( \lambda,\cU(\lambda),\cV(\lambda) )
                                                         \nonumber \\
  &&- \frac{k}{j+k} \cM(\lambda)^{j+k}
      b( \lambda,\cU(\lambda),\cV(\lambda) )
      \frac{\rd}{\rd \lambda}
      a( \lambda,\cU(\lambda),\cV(\lambda) ).
                                        \label{eq:extCommRelProof3}
\eeqnarray
The first term in the last three lines vanishes under the
$\lambda$-residue (because this is a total derivative with
respect to $\lambda$). The $\lambda$-derivatives in the other
terms can be evaluated by the chain rule and the second and
third equations of (\ref{eq:CanRel(lambda)}):
\beqnarray
      \frac{\rd}{\rd \lambda}
        a( \lambda,\cU(\lambda),\cV(\lambda) )
  &=& \left.\frac{\rd a(\lambda,u,v)}{\rd \lambda}
        \right|_{u=\cU(\lambda),v=\cV(\lambda)}
                                                         \nonumber \\
  &&  + \{ \cM(\lambda),
             a( \lambda,\cU(\lambda),\cV(\lambda) ) \}_\theta,
                                                         \nonumber \\
      \frac{\rd}{\rd \lambda}
        b( \lambda,\cU(\lambda),\cV(\lambda) )
  &=& \left.\frac{\rd b(\lambda,u,v)}{\rd \lambda}
        \right|_{u=\cU(\lambda),v=\cV(\lambda)}
                                                         \nonumber \\
  &&  + \{ \cM(\lambda),
             b( \lambda,\cU(\lambda),\cV(\lambda) ) \}_\theta.
                                          \label{eq:extCommRelProof4}
\eeqnarray
Plugging these formulas into the right hand side of
(\ref{eq:extCommRelProof3}), we obtain four terms originating
in the four terms on the right hand side of
(\ref{eq:extCommRelProof4}). We now separate them into two groups:
(a) those coming from the first terms on the right hand side of
(\ref{eq:extCommRelProof4}), and (b) those from the second terms
thereof.

(a) Contribution of these terms to (\ref{eq:extCommRelProof3})
can be written
\beqnarray
   &&  \left(
          \frac{j}{j+k} \cM(\lambda)^{j+k}
            a(\lambda,u,v)
              \frac{\rd b(\lambda,u,v)}{\rd \lambda}
       \right.
                                                      \nonumber \\
   &&  \left.\left.
        - \frac{k}{j+k} \cM(\lambda)^{j+k}
            b(\lambda,u,v)
              \frac{\rd a(\lambda,u,v)}{\rd \lambda}
                \right)
                  \right|_{u=\cU(\lambda),v=\cV(\lambda)}
                                                      \nonumber \\
   &=& - \int_0^{\cM(\lambda)} d\mu
           \left.
             \{ \cA(\lambda,\mu,u,v),
                \cB(\lambda,\mu,u,v) \}_{\lambda\mu}
           \right|_{u=\cU(\lambda),v=\cV(\lambda)}.
                                          \label{eq:extCommRelProof5}
\eeqnarray

(b) Contribution of these terms to (\ref{eq:extCommRelProof3})
is given by
\beqnarray
  &&   \frac{j}{j+k} \cM(\lambda)^{j+k}
       a( \lambda,\cU(\lambda),\cV(\lambda) )
         \{ \cM(\lambda),
           b( \lambda,\cU(\lambda),\cV(\lambda) ) \}_\theta
                                                      \nonumber \\
  && - \frac{k}{j+k} \cM(\lambda)^{j+k}
       b( \lambda,\cU(\lambda),\cV(\lambda) )
         \{ \cM(\lambda),
            a( \lambda,\cU(\lambda),\cV(\lambda) ) \}_\theta
                                                      \nonumber \\
  &=&  \frac{j}{j+k}
       a( \lambda,\cU(\lambda),\cV(\lambda) )
         \left\{ \dfrac{\cM(\lambda)^{j+k+1}}{j+k+1},
           b( \lambda,\cU(\lambda),\cV(\lambda) )
             \right\}_\theta
                                                      \nonumber \\
  && - \frac{k}{j+k}
       b( \lambda,\cU(\lambda),\cV(\lambda) )
         \left\{ \dfrac{\cM(\lambda)^{j+k+1}}{j+k+1},
           a( \lambda,\cU(\lambda),\cV(\lambda) )
             \right\}_\theta
                                                      \nonumber \\
  &=&  \left\{
         \frac{j}{(j+k)(j+k+1)}
           a( \lambda,\cU(\lambda),\cV(\lambda) )
             \cM(\lambda)^{j+k+1},
               b( \lambda,\cU(\lambda),\cV(\lambda) )
                  \right\}_\theta
                                                      \nonumber \\
  && - \left\{
         \frac{k}{(j+k)(j+k+1)}
           b( \lambda,\cU(\lambda),\cV(\lambda) )
             \cM(\lambda)^{j+k+1},
               a( \lambda,\cU(\lambda),\cV(\lambda) )
                  \right\}_\theta
                                                      \nonumber \\
  && - \{ a( \lambda,\cU(\lambda),\cV(\lambda) ),
          b( \lambda,\cU(\lambda),\cV(\lambda) ) \}_\theta
             \dfrac{\cM(\lambda)^{j+k+1}}{j+k+1}.
\eeqnarray
The first two terms in the last three lines vanish in
the $\theta$-integral (Lemma \ref{lem:vanishing}). The last term can
be rewritten
\beqnarray
  &&  - \left.
          \{ a(\lambda,u,v), b(\lambda,u,v) \}_{uv}
          \dfrac{\mu^{j+k+1}}{j+k+1}
        \right|_{\mu=\cM(\lambda),u=\cU(\lambda),v=\cV(\lambda)}
                                               \nonumber \\
  &=& - \int_0^{\cM(\lambda)} d\mu
          \left. \{ A(\lambda,\mu,u,v), B(\lambda,\mu,u,v) \}_{uv}
          \right|_{u=\cU(\lambda),v=\cV(\lambda)}.
                                          \label{eq:extCommRelProof6}
\eeqnarray

\noindent
Thus, contribution from each term of (\ref{eq:extCommRelProof3})
has been specified by (\ref{eq:extCommRelProof4}),
(\ref{eq:extCommRelProof5}) and (\ref{eq:extCommRelProof6}).
Inserting these pieces, the integral of (\ref{eq:extCommRelProof2})
can now be written
\beqnarray
  &&  \int \frac{d\theta_1 d\theta_2}{(2\pi)^2} \res_\lambda
          \cA( \lambda,\cM(\lambda),\cU(\lambda),\cV(\lambda) )
            \frac{\rd}{\rd \lambda}
              \cB( \lambda,\cM(\lambda),\cU(\lambda),\cV(\lambda) )
                                                      \nonumber \\
  &=& - \int \frac{d\theta_1 d\theta_2}{(2\pi)^2} \res_\lambda
          \int_0^{\cM(\lambda)} d\mu
            \{ \cA(\lambda,\mu,u,v),\cB(\lambda,\mu,u,v) \}
              |_{u=\cU(\lambda),v=\cV(\lambda)}
                                                      \nonumber \\
  &=& \delta_{ \{\cA,\cB\}_{\lambda\mu uv} } F.
\eeqnarray
Meanwhile, since $\cA(\lambda,0,u,v) = \cB(\lambda,0,u,v) = 0$,
the cocycle itself vanishes: $c(\cA,\cB) = 0$. Thus
(\ref{eq:extCommRelProof1}) turns out to be satisfied
in this case, too.

\section{Conclusion}

We have considered a new higher dimensional analogue of the
dispersionless KP hierarchy --- the toroidal model. The structure
of this model is almost parallel to the planar model In our
previous paper. These higher dimensional hierarchies have extra
spatial dimensions in addition to the two-dimensional ``phase
space'' variables $(k,x)$ of the dispersionless KP hierarchy.
In the toroidal model, they are compactified to a two (or any even)
dimensional torus.  Because of this, the Poisson algebra has a
``trace functional'' as mentioned in Introduction.  With this
linear functional, we have been able to introduce an analogue of
the $F$ function of the dispersionless KP hierarchy. We have also
constructed an infinite number of additional symmetries, and
shown that their commutation relations obey an underlying higher
dimensional Poisson algebraic structure. The $F$ function induces
an anomalous term in commutation relations of additional symmetries,
which strongly suggests that our definition of the $F$ function is
a natural generalization of the $F$ functions in lower dimensional
dispersionless integrable hierarchies.

Let us conclude this paper by pointing out two issues that should
be elucidated on the bases of our results.

1. {\it Field theoretic meaning of $\delta_\cA F$}.
In the case of the dispersionless KP hierarchy, we have a phase
space integral representation, (\ref{eq:dKPSymmFbis}), of
$\delta_\cA F$.  This expression is reminiscent of the ``fermi
fluid picture'' \cite{bib:fermi-fluid} of $c = 1$ matrix models.
Actually, they both stem from an underlying $1+1$ dimensional
free fermion system. In the case of (\ref{eq:dKPSymmFbis}), the
curve $\mu = \cM(\lambda)$ in the $(\lambda,\mu)$ space represent
a ``fermi surface'', and the curve $\mu = 0$ probably the bottom
of the ``fermi sea''.

Remarkably, we can rewrite the defining equation (\ref{eq:SymmF})
of $\delta_\cA$ in the toroidal model, too, can be rewritten
into a similar multiple integral:
\beqn
    \delta_\cA F
      = - \int \frac{d\theta_1 d\theta_2}{(2\pi)^2}
            \oint \frac{d\lambda}{2\pi i}
              \int_0^{\cM(\lambda)} d\mu
                \cA( \lambda,\mu,\cU(\lambda),\cV(\lambda) ).
                                         \label{eq:SymmFbis}
\eeqn
This poses an interesting question: What is a higher dimensional
counterpart of the $1+1$-dimensional free fermion theory?
This should be a kind of large-$N$ limit of the $N$-component
fermion system in the $N$-component KP hierarchy \cite{bib:DJKM},
but because of the ``multiplicative renormalization'' mentioned
in Introduction, the fermion system seems to turn into a
substantially distinct (stringy?) object.

2. {\it Moyal algebraic algebraic version of toroidal model}.
The planar model has a Moyal algebraic version \cite{bib:T-MoyalKP}.
It is not hard to construct a Moyal algebraic version of the
toroidal model. Since the toroidal Moyal algebra has a ``trace
functional'' as mentioned in Introduction, we have a chance to
construct a tau function of this hierarchy. If the tau function
defined by any means, the $F$ function will be characterized
as a leading term of $\hbar$-expansion
\beqn
    \tau(\hbar,t,x)
    = \exp\left( - \hbar^{-2} F(t,x) + O(\hbar^{-1}) \right).
                                         \label{eq:AsympTau}
\eeqn

This issue is also related to the previous one, i.e., searching
for a field theoretic representation of higher dimensional
integrable hierarchies.  It is well known that the tau functions
of the ordinary and multi-component KP hierarchies can be written
as an infinite determinant \cite{bib:Sato-Sato,bib:Segal-Wilson}.
The hypothetical tau (or $F$) function of our higher dimensional
hierarchy, too, might be a determinant on a huge vector space.
Actually, since the trace functional on a Moyal (or Poisson)
algebra is a multiplicatively renormalized large-$N$ limit of
the ordinary trace on $\gln(N)$, this ``determiant'' (unlike
the relatively naive infinite determinant for the KP hierarchy)
should be a considerably complicated quantity (a ``renormalized
determinant'', so to speak).

\end{document}